# Ryugu's observed volatile loss did not arise from impact heating alone

Kosuke Kurosawa[1], Ryota Moriwaki[2], Hikaru Yabuta[3], Ko Ishibashi[1], Goro Komatsu[4,1], and Takafumi Matsui[1,2]

[1]Planetary Exploration Research Center, Chiba Institute of Technology, 2-17-1, Tsudanuma, Narashino, Chiba 275-0016, Japan

[2]Institute of Geo-Cosmology, Chiba Institute of Technology, 2-17-1, Tsudanuma, Narashino, Chiba 275-0016, Japan

[3]Department of Earth and Planetary Systems Science, Graduate School of Science, Hiroshima University, 1-3-1, Kagamiyama, Higashi–Hiroshima, Hiroshima 739-8526, Japan

[4]International Research School of Planetary Sciences, Università d'Annunzio, Viale Pindaro, 42, 65127 Pescara, Italy

* Corresponding author: Kosuke Kurosawa (Kosuke.kurosawa@perc.it-chiba.ac.jp)

**Abstract**

Carbonaceous asteroids, including Ryugu and Bennu, which have been explored by the Hayabusa2 and OSIRIS-REx missions, were probably important carriers of volatiles to the inner Solar System. However, Ryugu has experienced significant volatile loss, possibly from hypervelocity impact heating. Here we present impact experiments at speeds comparable to those expected in the main asteroid belt (3.7 km s$^{-1}$ and 5.8 km s$^{-1}$) and with analogue target materials. We find that loss of volatiles from the target material due to impacts is not sufficient to account for the observed volatile depletion of Ryugu. We propose that mutual collisions in the main asteroid belt are unlikely to be solely responsible for the loss of volatiles from Ryugu or its parent body. Instead, we suggest that additional processes, for example associated with the diversity in mechanisms and timing of their formation, are necessary to account for the variable volatile contents of carbonaceous asteroids.

**Introduction**

Carbonaceous (C-type) asteroids are considered the parent bodies of carbonaceous chondrite meteorites because of their similar reflectance spectra[1,2]. For example, a typical CI chondrite (e.g., Orgueil) contains 19.9 wt.% water and 3.1 wt.% carbon[3]. Spectroscopic observations of asteroids show that the outer main asteroid belt consists mainly of C-type asteroids[4], which are considered to have been one of the major carriers of water and organic matter to the inner Solar System, including Earth[5]. Recently, two C-type asteroids, Ryugu and Bennu, were extensively investigated by Hayabusa2[6] and OSIRIS-Rex[7], respectively. The remote sensing observations revealed that Ryugu and Bennu have experienced numerous impacts at a variety of impact energies[8,9]. Ryugu appears to have lost a significant amount of its volatiles during its history as its surface only exhibits a weak OH absorption band at 2.72 μm[10]. In contrast, Bennu apparently did not experience such volatile loss[11]. Solar heating of Ryugu during its temporary orbital excursion near the Sun cannot explain its volatile depletion[12]. Two hypotheses have been proposed to explain the depletion in volatiles. The first is that the Ryugu parent body experienced global internal heating due to decay of short-lived radionuclides[8]. The second is that devolatilisation was due to catastrophic disruption of the Ryugu parent body[13]. Although other possibilities, such as inefficient aqueous alteration and heating due to multiple impacts, cannot be ruled out[8], the two aforementioned scenarios are the most realistic based on the remote sensing data and our current knowledge of meteorites and asteroids[8,13].

In this study, we address the feasibility of the impact scenario, which is one of the leading hypotheses to explain the "dry" nature of Ryugu and the difference between Ryugu and Bennu. Thus, if the shock response of C-type asteroids is accurately understood, then it may be possible to constrain the origins of Ryugu and Bennu. Previous impact experiments using metal plates flying at <2 km s$^{-1}$ [14–17], which is much slower than the typical impact velocity in the main belt region (4–5 km s$^{-1}$ [18]), showed that chondritic meteorites could have easily lost their volatile content during hypervelocity impacts. This is the reason why impact heating has been considered as the cause of the diversity in the volatile contents of carbonaceous meteorites and those observed in the two asteroids.

The legacy results, however, seem to contradict a recent spectroscopic survey[19] that showed a widespread survival of hydrous minerals across the main belt region despite the intense impact bombardment throughout its history. In this study, we investigated shock-driven volatile loss to assess whether hypervelocity impacts can cause significant volatile loss from C-type asteroids. Two-stage light gas guns allow the shock response of materials to be robustly investigated, because they can accelerate a macroscopic (>1 mm) spherical projectile up to speeds of several km s$^{-1}$ without heating of the projectile. However, chemical contamination from the gun operation has been a longstanding problem in such gas-gun experiments, including the 1980's experiments mentioned above. This has recently been overcome by the two-valve method[20]. In this study, we applied this experimental technique (Supplementary Note S1, Supplementary Table S1) to investigate the process of impact-driven volatile loss from carbonaceous chondrite-like materials (Supplementary Note S2, Supplementary Table S2), which are expected to be an analogue of C-type asteroid material.

**Results**

**Laboratory experiments of impact-driven volatile loss from chondrite-like materials**

We conducted hypervelocity impact experiments on a carbonaceous chondrite analogue with a mineral composition similar to the Orgueil meteorite[21] using a two-stage helium gas gun. The impact-generated gases were quantitatively measured with a quadrupole mass spectrometer (QMS), where the generated gas freely expands into a large free volume with respect to the volume of the shocked samples with volume ratios >10$^{6-7}$. In the experiments, Al$_2$O$_3$ projectiles were used that were accelerated to two different impact velocities of $v_{\text{imp}}$ = 3.7 km s$^{-1}$ (#459 and #461) and 5.8 km s$^{-1}$ (#440, #447, and #448). Hereafter, we refer to the two velocities as "low $v_{\text{imp}}$" and "high $v_{\text{imp}}$", respectively. Figure 1 shows the temporal variations of the ion currents of selected masses for a high $v_{\text{imp}}$ experiment (shot #448). The data supporting Fig. 1 can be found in Supplementary Data 1. The ion current for the mass number $M/Z = i$ is denoted as $I_i$. We confirmed that $I_{40}$ (Ar$^+$) was stable during the measurement. A rise in $I_4$ (He$^+$), which is the gas that propels the projectile, was not detected, suggesting that chemical contamination was negligible. Abrupt increases in $I_2$ ($H_2^+$), $I_{16}$ ($CH_4^+$), $I_{28}$ ($CO^+$), $I_{44}$ ($CO_2^+$), and $I_{34}$ ($H_2S^+$) were detected. In contrast, we did not detect a sudden increase in $I_{18}$ ($H_2O^+$), although our experimental system is capable of detecting impact-generated water vapour from hydrous minerals [Kurosawa et al., 2019]. $I_{18}$ did not spike immediately after the impact, but gradually increased with time. The time profile for $I_{18}$ is different from the other species, suggesting that water vapour in micropores in the porous target was released due to the destruction of the target after the impact. For the other two shots at high $v_{imp}$ (#440 and #447), the results are qualitatively the same as #448, although the peak values of $I_i$ slightly differ. At low $v_{imp}$ (#459 and #461), the peak $I_i$ values for the species were several times lower than for the high $v_{imp}$ experiments and a rise in $I_{34}$ ($H_2S^+$) was not observed. We also conducted a series of calibration experiments (Supplementary Notes S3 and S4, Supplementary Figures S1–S9, Supplementary Table S3). The contributions to the ion currents for lighter species from cracking of heavier species due to electron impacts in the QMS were investigated using standard gases. We determined such contributions to $I_i$ from the raw signals by using the cracking patterns obtained for standard gases. This procedure is described in the Supplementary Note S4. We measured the amounts of the detected species quantitatively by integrating the time profiles of the ion currents and assuming the ratios of the integrated values to that of $CO_2$ were linearly proportional to the molar ratios to $CO_2$. We confirmed that this assumption is valid, at least for the molar ratio of $CH_4$ to $CO_2$, in the calibration experiments (Supplementary Note S3). Figure 2 shows the produced micromoles of $H_2$, $CH_4$, $CO$, $CO_2$, and $H_2S$. The main product was $CO_2$ in all shots. The mass of the $CO_2$ production corresponds to 0.2–0.3 and 1–2 wt.% of the projectile mass $M_p$ at low and high $v_{imp}$, respectively. The amounts of the reducing species ($H_2$, $CH_2$, and $CO$) that were generated were 10–50 mol.% of the $CO_2$ produced. The amount of sulphur-bearing gases produced was only 0.01–0.1 mol.% of the carbon-bearing gases, although the molar ratio of sulphur to carbon of the simulant is close to unity. We also found that water loss from hydrous minerals in the target (~50 wt.% of the carbonaceous chondrite analogue) was not significant. The upper bound of the released water vapour was estimated to be 7 wt.% of $M_p$ (Supplementary Note S3, Supplementary Figure S5). We confirmed that the chemical composition (e.g., $CO/CO_2$ and $H_2/CO$) of the generated species does not vary with $v_{imp}$. In contrast, the production of $H_2$, $CH_4$, $CO$, and $CO_2$ depends strongly on $v_{imp}$ ($\propto v_{imp}^{3.5}$). We also conducted a control experiment with a basalt block, which is volatile-free, and confirmed that the detected species did indeed originate from the simulated CI meteorite target (Supplementary Note S5, Supplementary Figures S10 and S11).

**Discussion**
**Hypervelocity impacts in the main asteroid belt are not responsible for volatile loss**

The pressures and temperatures experienced in the target were estimated with the iSALE-2D shock physics code[22–24] (Supplementary Notes S6 and S7, Supplementary Tables S4–S6). We used the ANalytical Equations Of State (ANOES)[25] to describe the thermodynamic response of a simplified model target composed of 25% porous serpentine. The residual (post-shock) temperatures of the heated material were derived from the simulations (i.e., the temperature after release from a state of high shock pressure). Note that this is in contract to the results reported in Michel et al. (2020)[13], in which peak-shock temperatures were considered. This allowed us to accurately estimate the role of impact heating on impact-induced devolatilisation.

We conducted two sets of numerical simulations called the Lab–simulation and Ryugu–simulation (see Methods). Figure 3 shows typical examples of the simulations, which plot the expected residual temperature after a pressure release $T_{\text{res}}$ (Supplementary Note S8, Supplementary Figure S12, Supplementary Table S7). The region highlighted in red indicates the heated region that experienced partial devolatilisation. Both simulations exhibit a similar thermal structure such that highly heated materials ($T_{\text{res}} \sim 800$ K where devolatilisation of the CI chondrite analogue initiates) are initially located within a distance of 1–2 $D_p$ from the impact point and form a thin high-temperature layer beneath the crater floor. Devolatilisation can only occur from this layer. Figure 4 shows the cumulative mass as a function of peak pressure and entropy. The data supporting Figs. 4a and 4b can be found in Supplementary Data 2 and 3, respectively. The corresponding impact velocities required to produce a similar mass distribution in the case of a hypothesised catastrophic collision onto the parent body of Ryugu pertaining to low and high $v_{\text{imp}}$ are estimated to be 4.0–4.5 and 6.0–7.0 km s$^{-1}$, respectively, in terms of both the peak pressure and expected residual temperature. Given that the most probable $v_{\text{imp}}$ in the main asteroid belt region is estimated as 4.4 km/s[18], our impact experiments at low and high $v_{\text{imp}}$ simulate typical and relatively high-speed collisions, respectively. It is worthwhile to mention here on the results by Tyburczy et al. (1986)[17]. They conducted shock recovery experiments on chips of the Murchison meteorite, and achieved relatively high pressures as shown in the Fig. 4a. They also detected volatile loss from the target. However, their experiment was conducted in a closed system where a sample was sealed in a stainless-steel container. Thus, the thermodynamic path experienced by the target was rather complex, and it is difficult to apply these results to natural impact events[26]. The details of the shock physics modelling are summarised in the Supplementary Notes S6 and S7.

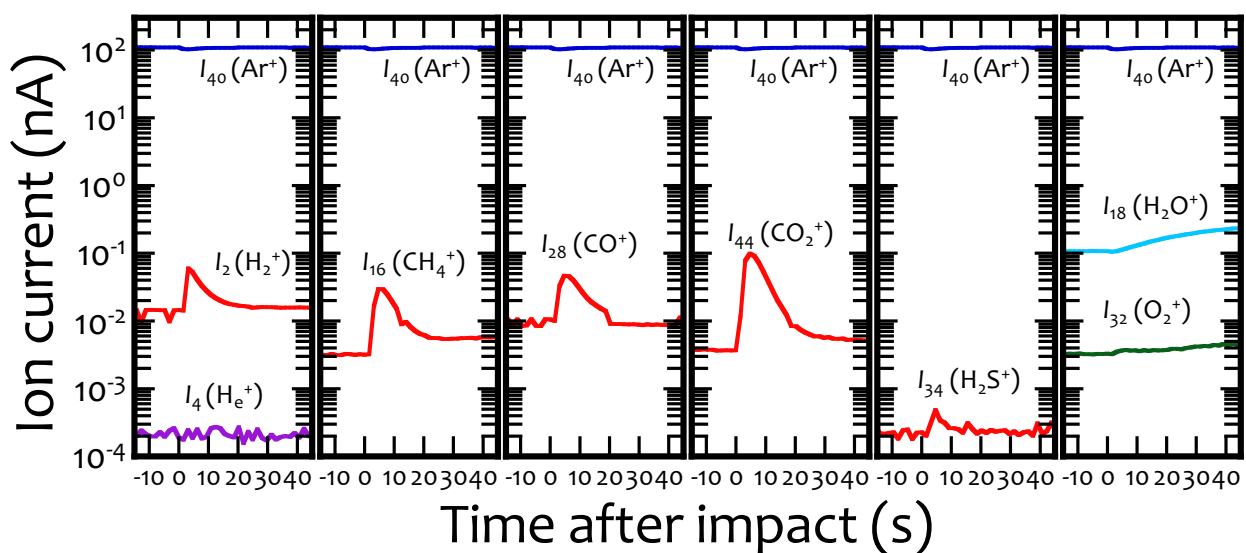

**Figure 1. Mass spectrometry data for the generated gas species.** The graphs show the temporal variation of the ion currents for gases with the mass numbers $I_i$ for shot #448 (high $v_{imp}$). The time variations for the ion counts can be found in Supplementary Data 1.

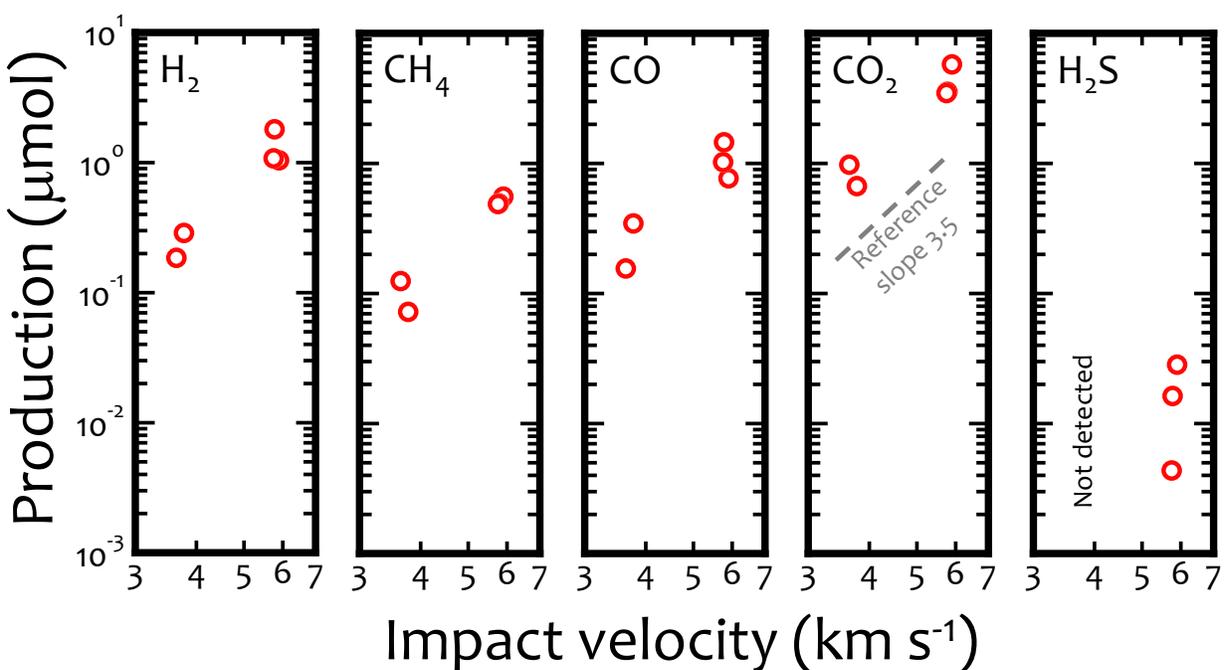

**Figure 2. Plots of the amounts of impact-produced species vs. impact velocity.** Note that both the *x*- and *y*-axes are shown with a log scale. In the panel for $CO_2$, a reference slope that represents a power law with an exponent of 3.5 is also shown. The data supporting Fig. 2 are summarised in the Supplementary Table S2.

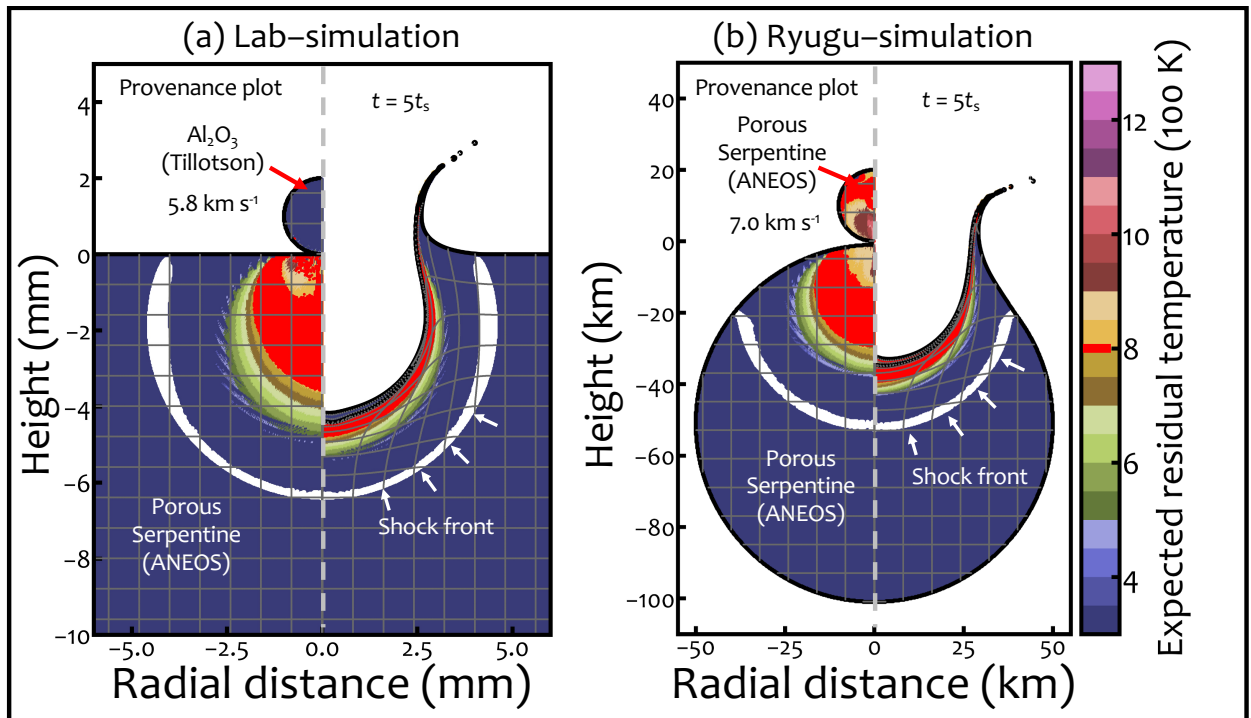

**Figure 3. Numerical results. (a)** The result of the Lab-simulation at the high $v_{\text{imp}}$ obtained with the iSALE shock physics code. A snap shot (right) and provenance plot (left) are shown. The colour indicates the expected residual temperature $T_{\text{res}}$ calculated from the temporal entropy. A grid of tracer probes, plotted every 25 cells in the horizontal and vertical directions, is shown as grey lines. The location of the shock front is shown in a white line. Note that we did not calculate $T_{\text{res}}$ in the Al$_2$O$_3$ projectile. **(b)** The same as panel (a), except the result of the Ryugu-simulation is shown. In this calculation, the impact velocity was set to 7 km s$^{-1}$. In these panels, the results are shown for $t = 5t_s$ ($t$ is the time after impact), where $t_s = D_p/v_{\text{imp}}$ and is the characteristic time for projectile penetration, and $D_p$ is the projectile diameter. This time was chosen because at this point the impact heating is finished.

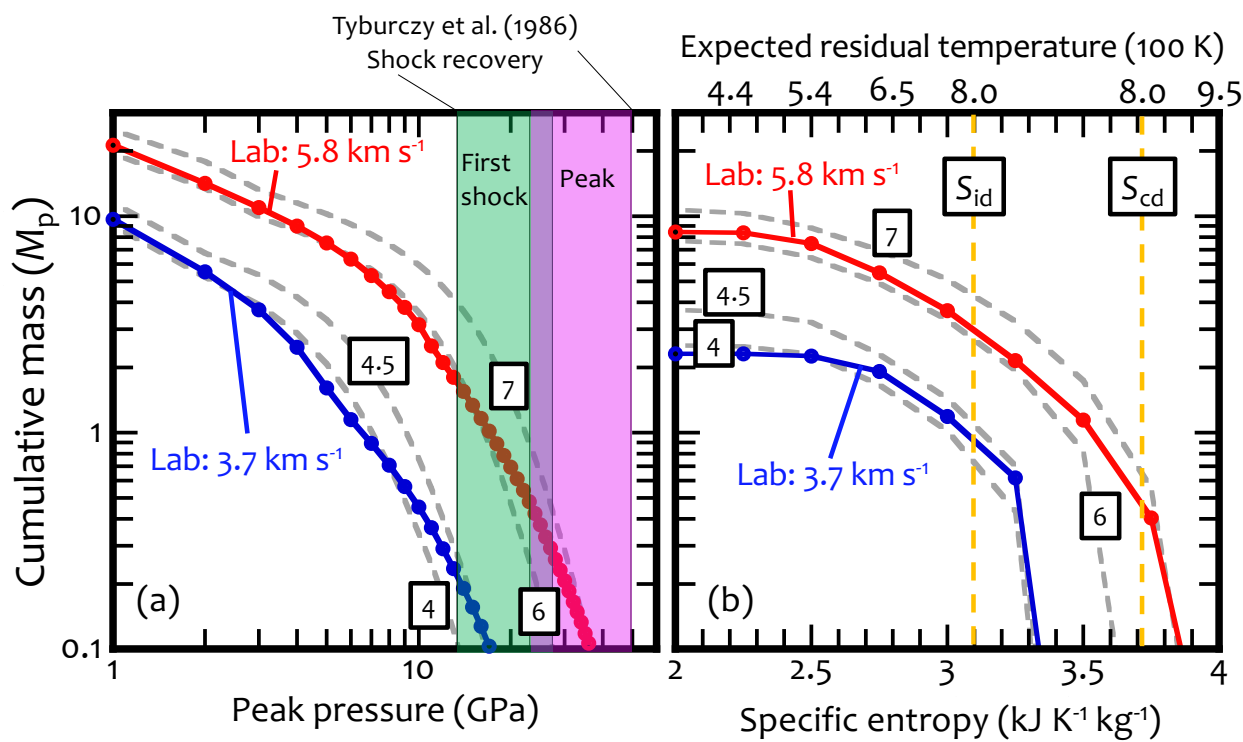

**Figure 4. Pressure and entropy (post-shock residual temperature) experienced during impacts. (a)** The cumulative mass at a given peak pressure as a function of the peak pressure. The blue and red points correspond to the results of the Lab–simulation at low and high $v_{imp}$, respectively. The four dashed grey lines are the results of the Ryugu–simulation. The numbers in the black boxes on the lines are the impact velocities used in the Ryugu–simulation. The green and pink shaded regions indicate the range of pressures in a previous impact experiment[17]. **(b)** The same as panel (a), except the cumulative mass at a given entropy is shown as a function of entropy. The vertical dashed yellow lines indicates the entropies required for incipient and complete devolatilisation of the target, respectively (Supplementary Note S8). The expected residual temperature $T_{res}$ is also shown on the top $x$-axis. The calculated results shown in Figs. 4a and 4b can be found in Supplementary Data 2 and 3, respectively.

Here we discuss the nature of impact devolatilisation of the simulated CI meteorite material. By detailed consideration of the volatile budget in the target material, energy partitioning during shock propagation, and the chemical composition of the generated gas (Supplementary Note S8, Supplementary Figures S13–S16, Supplementary Table S8), we infer that the devolatilised volume is limited to only ~10% of the predicted value due to local energy concentration regardless of impact velocity. The temperature of the local hot region is estimated to be 1200–1650 K based on the measured $CO/CO_2$ molar ratio and oxygen fugacity assuming a CI composition[27]. The detected

$H_2$, CO, and $CO_2$ are expected to have been produced by oxidation of the organics by water vapour and molecular oxygen derived from dehydration of Mg-serpentine and thermal breakdown of magnetite into wüstite and/or metallic iron in the vicinity of the local hot region. Mg-serpentine and magnetite are the first- and second-most abundant minerals in the CI meteorite analogue. The local hot region is likely to be produced by the shear velocities between different grains in the target, because of the large density contrast between the constituent minerals (1–5 Mg m$^{-3}$) (Supplementary Note S8, Supplementary Figure S15). The released water vapour from the local hot region is estimated to be 1–2 wt% (Supplementary Note S8, Supplementary Figure S16). The temperature outside of the hot region would not have been high enough to cause oxidation of the organics. Outside the local hot region, only pure pyrolysis of the organics in the target occurred (700–800 K[21]). The measured $CH_4$ would have come from this region.

We now consider the effect of the impact obliquity on the total gas production. Although we performed vertical impact experiments in our laboratory, virtually all natural impacts occur at oblique incident angles to target bodies[28]. A previous hydrocode simulation[29], which did not consider any material strengths in their calculations, shows that the degree of shock heating decreased with the lowering impact angle measured from the target surface. In contrast, an oblique impact experiment[30] demonstrated that subsequent shear heating after shock propagation rather enhanced the total gas production during oblique impacts. A three-dimensional shock physics modelling[31] confirmed the role of the shear heating, but also showed that the degree of impact heating due to both shock and post-shock shear heating in oblique impacts of 45 degrees was similar to that pertaining to vertical impacts. Thus, it seems that the enhancement of shear heating during oblique impacts acts to compensate for the weaker shock heating during oblique impacts than vertical impacts. Consequently, we suggest that the total gas production obtained in this study would be a good approximation in the case of oblique impacts. Nevertheless, future experimental confirmation is required on this point.

Assuming that the simulated CI meteorite material is a suitable analogue of the constituent materials of C-type asteroids, our results predict the response of C-type asteroids to impact shocks as follows: (1) impact-induced volatile release, including water vapour and C-bearing species, is limited to only 2–4 wt.% of the impactor mass, even for a relatively high-speed collision in the main asteroid belt (6–7 km s$^{-1}$). This inefficient devolatilisation occurs because the devolatilised volume is limited by the local energy concentration due to frictional heating between the different grains; (2) devolatilisation of S-bearing minerals during a typical collision in the main asteroid belt is negligible, resulting in sulphur enrichment in the impactor footprint. Our results are consistent with two seemingly contradictory facts: the ubiquitous presence of hydrous minerals on C-type asteroids in the main belt[19] and non-negligible amounts of thermally metamorphosed (up to ~1000 K) CI/CM chondrites[32–34]. Consequently, the volatile depletion that is observed on the entire surface of Ryugu is unlikely to be due to a single catastrophic collision[13] or multiple impacts[8], at least at mutual collision velocities of up to 7 km s$^{-1}$, regardless of impactor density (Supplementary Note S9, Supplementary Figure S17, Supplementary Tables S9 and S10). It should be mentioned here that we cannot rule out the possibility that massive volatile loss on Ryugu or its parent body was caused by extremely high-speed collisions (Supplementary Note S10, Supplementary Figure S18) because an intensely heated volume experiencing complete devolatilization becomes a significant fraction at >7 km s$^{-1}$. Although such collisions are statistically rare (<15%)[18], impact devolatilization of carbonaceous asteroids during high-speed collisions at >7 km s$^{-1}$ should be explored with future impact experiments in a fully open system. Finally, we propose that the bulk molar ratio of sulphur to carbon in returned samples may be a useful tracer of hypervelocity impacts onto Ryugu and Bennu and their parent bodies.

## Methods
### Impact experiments

The impact experiments were conducted using a two-stage light gas gun in the Hypervelocity Impact Facility of the Planetary Exploration Research Center of the Chiba Institute of Technology (PERC/Chitech), Japan[35]. In this study, we focused on the shock responses of the constituent materials of C-type asteroids. However, C-type asteroid material is not yet available. In this study, C-type asteroid simulant provided by Exolith (https://sciences.ucf.edu/class/exolithlab/) was used as the target (Supplementary Note S2). We made pellets of this material with a hydraulic pump that were 54 mm in diameter and 30 mm thick. The bulk density of the pellets is $1.92 \pm 0.03$ Mg m$^{-3}$, which is close to the typical density of carbonaceous chondrites[36]. The target porosities were calculated with the dimensions and mass of the target pellets and a typical grain density (2.5 Mg m$^{-3}$) of carbonaceous chondrite meteorites[36]. Given that the shape of the pellets is cylindrical, we can easily measure the radius and length of the pellets and calculate the target volumes. The porosity was estimated to be 22%–24%.

In order to investigate the devolatilisation behaviour of the target, we used the two-valve method developed in a previous study[20]. This method allows us to measure shock-generated gas in an open system, which has the same geometry as natural impacts and low levels of chemical contamination from the gun operation. In addition, we used helium gas instead of the normally used hydrogen gas for the projectile acceleration, in order to exclude hydrogen from the experimental chamber as this contributes to chemical reduction. A quadrupole mass spectrometer (QMS; Pfeiffer Vacuum Technology AG; Prisma plus QMG220) was used to analyse the shock-generated gas after each impact. An Al$_2$O$_3$ sphere that was 2 mm in diameter was used as the projectile, which was accelerated using a nylon slit sabot[37]. In this study, we conducted five shots (#440, #447, #448, #459, and #461) at two different impact velocities of $v_{imp}$ = 3.7 km s$^{-1}$ (#459 and #461) and 5.8 km s$^{-1}$ (#440, #447, and #448) as described in the main text. Further details of the impact experiments, including the experimental procedures, the procedure of data reduction, and impact velocities achieved, are described in the Supplementary Note S1.

We also conducted a series of calibration experiments using a gas injection device and gas mixtures with different ratios of CO$_2$ to Ar, because CO$_2$ was the dominant gas detected after all the shots. A gas mixture of Ar and CO$_2$ was injected into the experimental chamber for a short duration (~0.1 s) using an automatic gate valve. Further details of the calibration are described in the Supplementary Note S3.

### Numerical simulations

The ANalytical Equation Of State (ANEOS)[25] for serpentine[38] was used in the simulations. Serpentine is the main constituent of the simulated CI meteorite material and possibly of carbonaceous asteroids. The $\varepsilon$–$\alpha$ porosity compaction model[24] was also used to account for the porosity. We set the initial porosity to 25% to match the porosity of our target. This value is similar to that of carbonaceous chondrites[36].

The Tillotson equation of state[39] for Al$_2$O$_3$[40] was used for the projectile in the Lab–simulation that simulated the laboratory impact experiments. The impact velocity was set to be the same as that achieved in the experiments. Given that the projectile is denser (4.0 Mg m$^{-3}$) than the porous target (1.9 Mg m$^{-3}$), our laboratory impact experiments simulated impacts between two porous asteroids at a somewhat higher $v_{imp}$ than the velocity achieved in the experiments, in terms of the experienced pressure and temperature in the target body.

In the Ryugu–simulation, we modelled vertical impacts of a spherical impactor with a diameter of 20 km onto a hypothetical Ryugu parent body with a 100 km diameter (Fig. 3b). Note that the size of the parent body is still debated. The calculation settings are similar to those used in a previous study[41], except that both the impactor and target were assumed to be porous serpentine with a porosity of 25%. The specific impact energy $Q$ was varied from 1 to 400 kJ kg$^{-1}$, which is close to the range of the required energy densities between cratering and catastrophic disruption $Q^*$ of 100-km-sized bodies[42,43]. The parameter sets and calculation settings are summarised in the Supplementary Notes S6 and S7. Our experimental results also help to predict the outcomes of impact events onto CI chondrites at any scale, although we considered a large-scale impact event in this study.

**Data Availability:**
The data supporting Figures 1 and 4 can be available as Supplementary data available at https://zenodo.org/record/5055018#.YN239y33J4Q. The data supporting Figure 2 is listed in Supplementary Table S3. The calculation results shown in Figure 3 can be reproduced with the input parameters listed in Supplementary Table S4–S6.

**Code Availability**
The iSALE shock physics code is not fully open-source, and it is distributed on a case-by-case basis to academic users in the impact community for non-commercial use only. A description of the application requirements can be found at the iSALE website (http://www.isale-code.de/redmine/projects/isale/wiki/Terms_of_use).

**Acknowledgements:** This work was supported by ISAS/JAXA as a collaborative program with the Hypervelocity Impact Facility. We appreciate Sunao Hasegawa and Takaya Okamoto for assistance with the development of our experimental system. We thank Dimitri Veras and Tom Davison for their helpful comments about English writing. We also thank the developers of iSALE, including G. Collins, K. Wünnemann, B. Ivanov, J. Melosh, D. Elbeshausen, and Tom Davison. We thank three referees for their careful reviews that helped greatly improve the manuscript, and the editor for handling the manuscript. A part of numerical computations was carried out on the general-purpose PC cluster at Center for Computational Astrophysics, National Astronomical Observatory of Japan. K.K. is supported by JSPS KAKENHI grants JP17H01176, JP17H01175, JP17K18812, JP17H02990, JP18H04464, and JP19H00726, and by the Astrobiology Center of the National Institute of Natural Sciences (Grants AB261014 and AB281026).

**Author Contributions**
K.K. carried out the impact experiments, analyses, and numerical simulations, and wrote the initial draft of the manuscript. K.K., G.K., H.Y., and T.M. conceived the initial research idea and designed the impact experiments. R.M. and K.I. assisted with the impact experiments and the interpretation of the results. All the authors discussed the results and their implications.

**Competing Interests Statement**
The authors declare no competing interests.

# References

1. Johnson, T. V. & Fanale, F. P. Optical properties of carbonaceous chondrites and their relationship to asteroids. *Journal of Geophysical Research* **78**, 8507–8518 (1973).
2. Hiroi, T., Zolensky, M. E., Pieters, C. M., and Lipschutz, M. E. Thermal metamorphism of the C, G, B, and F asteroids seen from the 0.7 um, 3 um, and UV absorption strengths in comparison with carbonaceous chondrites. *Meteoritics & Planetary Science* **31**, 321–327 (1996).
3. Baker, B. L. Review of organic matter in the Orgueil meteorite. *Space Life Sciences* **2**, 472–497 (1971).
4. Bus, J. S. & Binzel R. P. Phase II of the small main-belt asteroid spectroscopic survey: A feature based taxonomy. *Icarus* **158**, 146-177.
5. Hayatsu, R. & Anders, E. Organic compounds in meteorites and their origin. *Top Curr. Chem.* **99**, *1c-37* (1981).
6. Watanabe, S. et al. Hayabbusa2 arrives at the carbonaceous asteroid 162173 Ryugu - A spinning top-shaped rubble pile. *Science* **364**, 268–272 (2019).
7. Lauretta, D. S. et al. The unexpected surface of asteroid (101955) Bennu. *Nature* **568**, 55–60 (2019).
8. Sugita, S. et al. The geomorphology, color, and thermal properties of Ryugu: Implications for parent-body processes. *Science* **364**, eaaw0422 (2019).
9. Walsh, K. J. et al. Craters, boulders and regolith of (101955) Bennu indicative of an old and dynamic surface. *Nature Geoscience* **12**, 242–246 (2019).
10. Kitazato, K. et al. The surface composition of asteroid 162173 Ryugu from Hayabusa2 near-infrared spectroscopy, *Science* **364**, 272–275 (2019).
11. Hamilton, V. E. et al. Evidence for widespread hydrated minerals on asteroid (101955) Bennu. *Nature Astronomy* **3**, 332–340 (2019).
12. Morota, T. et al. Sample collection from asteroid 162173 Ryugu by Hayabusa2: implications for surface evolution. *Science* **368**, 654–659 (2020).
13. Michel, P. et al. Collisional formation of top-shaped asteroids and implications for the origins of Ryugu and Bennu. *Nature Communications* **11**, 2655 (2020).
14. Boslough, M. B., Weldin, R. J., & Ahrens, T. J. Impact-induced water loss from serpentine, nontronite and kernite, Proc 11[th] Lunar Scie Conf., pp. 2145-2158, (1980).
15. Lange, M. A. & Ahrens, T. J. Impact-induced dehydration of serpentine and the evolution of planetary atmospheres, *Proc. 13[th] Lunar Planet. Sci. Conf., Part1, Journal of Geophysical Research* **87**, A451-A456, (1982).
16. Lange, M. A., Lambert, P., & Ahrens, T. J. Shock effects on hydrous minerals and implications for carbonaceous meteorites, *Geochim. Cosmochim. Acta* **49**, 1715-1726, (1985).
17. Tyburczy, J. A., Frisch, B., and Ahrens, T. J. Shock-induced volatile loss from a carbonaceous chondrite: implications for planetary accretion. *Earth and Planetary Science Letters* **80**, 201–207 (1986).
18. Bottke, W. F., Nolan, M. C., Greenberg, R., & Kolvoord, R. A. Velocity distributions among colliding asteroids, *Icarus* **107**, 255–268 (1994).
19. Usui, F., Hasegawa, S., Ootsubo, T., and Onaka, T. AKARI/IRC near-infrared asteroid spectroscopic survey: AcuA-spec. *Publ. Astron. Soc. Japan* **71**, 1-41, (2019)
20. Kurosawa, K., Moriwaki, R., Komatsu, G., Okamoto, T., Sakuma, H., Yabuta, H. & Matsui, T. Shock vaporization/devolatilization of evaporitic minerals, halite and gypsum, in an open system investigated by a two-stage light gas gun. *Geophysical Research Letters* **46**, 7258–7267 (2019).
21. Britt, D. T., Cannon, K. M., Donaldson Hanna K., Hogancamp, J., Poch, O., Beck, P., Martin, D., Escrig, J., Bonal, L., & Metzger, P. T. Simulated asteroid materials based on carbonaceous chondrite mineralogies. *Meteoritics & Planetary Science* **54**, 2067–2082 (2019).
22. Amsden, A., Ruppel, H. & C. Hirt. SALE: A simplified ALE computer program for fluid flow at all speeds. *Los Alamos National Laboratories Report*, LA-8095:101p (1980).
23. Ivanov, B. A., Deniem, D. & Neukum G. Implementation of dynamic strength models into 2-D hydrocodes: Applications for atmospheric breakup and impact cratering. *Int. J. Impact Eng.* **20**, 411–430 (1997).
24. Wünnemann, K., Collins, G. S. & Melosh, H. J. A strain-based porosity model for use in hydrocode simulations of impacts and implications for transient crater growth in porous targets. *Icarus* **180**, 514–527 (2006).
25. Thompson, S. L. & Lauson, H. S. Improvements in the Chart-D radiation hydrodynamic code III: Revised analytical equation of state, pp. *SC-RR-71 0714* **119** pp., Sandia Laboratories, Albuquerque, NM (1972).
26. Ivanov, B. A. & Deutsch, A. The phase diagram of CaCO3 in relation to shock compression and decompression. *Physics of the Earth and Planetary Interiors* **129**, 131-143, (2002).
27. Schaefer, L., & Fegley, B. Jr. Redox states of initial atmospheres outgassed on rocky planets and planetesimals. *The Astrophysical Journal* **843**, 120 (2017).
28. Shoemaker, E. M. Impact mechanics at Meteor Crater, Arizona. In B. M. Middlehurst, & G. P. Kuiper (Eds.), The solar system, (Vol. 4, pp. 301–336). Chicago, IL: Univ. of Chicago Press, (1963).
29. Pierazzo, E. & Melosh, H. J. Melt production in oblique impacts. *Icarus* **145**, 252-261, (2000).
30. Schultz, P. H. Effect of impact angle on vaporization. *Journal of Geophysical Research* **100**, 21117–21135, (1996).
31. Wakita, S., Genda, H., Kurosawa, K., & Davison, T. M. Enhancement of impact heating in pressure-strengthened rocks in oblique impacts. *Geophysical Research Letters* **46**, (2019)
32. Yabuta, H., Alexander, C. M. O'D., Fogel, M. L., Kilcoyne, A. L. D., & Cody, G. D. A molecular and isotopic study of the macromolecular organic matter of the ungrouped C2 WIS 91600 and its relationship to Tagish Lake and PCA 91008. *Meteoritics & Planetary Science* **45**, 1446–1460 (2010).
33. King, A. J., Solomon, J. R., Schofield, P. F., & Russell, S. S. Characterising the CI and CI-like carbonaceous chondrites using thermogravimetric analysis and infrared spectroscopy. Earth, Planets and Space **67**, 198 (2015).
34. King, A. J., Schofield, P. F., & Russell, S. S. Thermal alteration of CM carbonaceous chondrites: Mineralogical changes and metamorphic temperatures. *Geochimica et Cosmochimica Acta* **298**, 167–190, (2021).
35. Kurosawa, K., Nagaoka, Y., Senshu, H., Wada, K., Hasegawa, S., Sugita, S. & Matsui, T. Dynamics of hypervelocity jetting during oblique impacts of spherical projectiles investigated via ultrafast imaging. *Journal of Geophysical Research Planets* **120**, 1237–1251 (2015).
36. Ostrowski, D. and Bryson, K. The physical properties of meteorites. *Planetary and Space Science* **165**, 148–178 (2019).
37. Kawai, N., Tsurui, K., Hasegawa, S. & Sato, E. Single microparticle launching method using two-stage light gas gun for simulating hypervelocity impacts of micrometeoroids and space debris. *Reviews of Scientific Instruments* **81**, 115105 (2010).

38. Brookshaw, L. An Equation of state for serpentine, *Tech. Rep., working paper series* SC-MC-9813 (Queensland: Faculty of Sciences, University of Southern Queensland) (1998).
39. Tillotson, J. H. Metallic equations of state for hypervelocity impact. *Technical Report GA-3216*, General Atomic Report (1962).
40. Kurosawa, K., Genda, H., Azuma, S., & Okazaki, K. The role of post-shock heating by plastic deformation during impact devolatilization of calcite ($CaCO_3$). *Geophysical Research Letters* **48**, e2020GL091130. https://doi.org/10.1029/2020GL091130 (2021).
41. Wakita, S. and Genda, H. Fates of hydrous materials during planetesimal collisions. *Icarus* **328**, 58–68 (2019).
42. Jutzi, M., Michel, P., Benz, W., and Richardson, D. C. Fragment properties at the catastrophic disruption threshold: the effect of the parent body internal structure. *Icarus* **207**, 54–65, (2010).
43. Suetsugu, R., Tanaka, H., Kobayashi, H., and Genda, H. Collisional disruption of planetesimals in the gravity regime with iSALE code: Comparison with SPH code for purely hydrodynamic bodies. *Icarus* **314**, 121–132 (2018).

**Supplementary Information**

**<u>Ryugu's observed volatile loss did not arise from impact heating alone</u>**

Kosuke Kurosawa[1], Ryota Moriwaki[2], Hikaru Yabuta[3], Ko Ishibashi[1], Goro Komatsu[4,1], and Takafumi Matsui[1,2]

[1]Planetary Exploration Research Center, Chiba Institute of Technology, 2-17-1, Tsudanuma, Narashino, Chiba 275-0016, Japan

[2]Institute of Geo-Cosmology, Chiba Institute of Technology, 2-17-1, Tsudanuma, Narashino, Chiba 275-0016, Japan

[3]Department of Earth and Planetary Systems Science, Graduate School of Science, Hiroshima University, 1-3-1, Kagamiyama, Higashi–Hiroshima, Hiroshima 739-8526, Japan

[4]International Research School of Planetary Sciences, Università d'Annunzio, Viale Pindaro, 42, 65127 Pescara, Italy

* Corresponding author: Kosuke Kurosawa (<u>Kosuke.kurosawa@perc.it-chiba.ac.jp</u>)

**Contents of this file**

Supplementary Notes        S1–S10
Supplementary Figures      S1–S18
Supplementary Tables       S1–S10

**Introduction**

   This document includes: (1) a detailed description of the impact experiments (Supplementary Note S1; Supplementary Table S1); (2) a description of the samples used in the experiment (Supplementary Note S2; Supplementary Table S2); (3) the methods and results of the calibration experiment and total amounts produced of the detected species (Supplementary Note S3; Supplementary Figs S1–S5; Supplementary Table S3); (4) the data reduction procedures (Supplementary Note S4; Supplementary Figs S6–S9); (5) a comparison with a control experiment (Supplementary Note S5; Supplementary Figs S10–S11); (6) the input parameters and calculation settings of the iSALE computations (Supplementary Note S6; Supplementary Tables S4–S6); (7) the impact velocities in the main asteroid belt region (Supplementary Note S7); (8) a discussion on the mechanism of impact devolatilisation of CI-like materials (Supplementary Note S8; Supplementary Figs S12–S16; Supplementary Tables S7–S8); (9) the effects of impactor density (Supplementary Note S9; Supplementary Fig. S17; Supplementary Tables S9–S10) and (10) the applicable limit of Applicable limit of the experimental results in terms of impact velocity (Supplementary Note S10; Supplementary Fig. S18).

**Supplementary Note S1. Hypervelocity impact experiments**

A two-stage light gas gun in the Hypervelocity Impact Facility of the Planetary Exploration Research Center of the Chiba Institute of Technology (PERC/Chitech), Japan[1], was used in this study. The two-valve method[2] was used to avoid chemical contamination from the gun operation, as mentioned in the main text. A schematic diagram and the detailed operating conditions of the two-valve method can be found in Kurosawa et al. (2019)[2]. Here, we briefly describe the experimental procedures and how these differ from the original study. We installed two automatic gate valves (#1 and #2) between the gun and experimental chamber where the target is placed and impacts occur. The gate valves #1 and #2 were inserted into the upper- and down-range of the gun system, respectively. The gun side up-range of valve #1 is evacuated and down-range from valve #1 is filled with a chemically inert gas to 500 Pa. The volume filled with the inert gas is then gently evacuated with a rotary pump. To maintain a constant pressure in the experimental chamber, the inert gas was continuously supplied into the experimental chamber at the same gas flux as the evacuation. As a result, the inert gas flows from the experimental chamber to a quadrupole mass spectrometer (QMS; Pfeiffer Vacuum Technology AG; Prisma plus QMG220) was produced. We operated the gun and two gate valves using specific time delays with a pulse generator. At the beginning of the experiment, the valve #1 is opened, resulting in outflow of inert gas to the gun side. The gun is then operated and a projectile is accelerated towards the target placed in the experimental chamber. The contaminant gas produced by the gun operation following the projectile cannot intrude into the experimental chamber due to the outflow. In contrast, the projectile can easily penetrate into the outflow and collide with the target. The valve #2 is closed immediately after the impact to retain the gases produced by impact-induced devolatilisation of the target. Finally, the generated gas was introduced into the QMS by the inert gas flow. Consequently, we are able to measure the chemical composition of the gas generated by an impact under the same geometry conditions as natural impacts, with a small risk of chemical contamination from the gun operation.

The remaining issue with the two-valve method is quantification of the generated gas due to gas escape from the experimental chamber prior to the shutdown of the valve #2. To overcome this, we changed the carrier gas from helium to argon, and inserted an extension tube between the valve #2 and the experimental chamber. Given that the speed of sound $C_s$ in argon is three times slower than that in helium at the same temperature ($C_s = \sqrt{\frac{\gamma RT}{\mu}}$, where $\gamma$, $R$, $T$, and $\mu$ are the ratio of specific heat, gas constant, temperature, and atomic weight, respectively), the escape gas flux in the case of Ar is much smaller than for helium. We conducted the same test as carried out in the original study (see Text S3.5 in the Supplementary Information of Kurosawa et al. [2019][2]) to examine the degree of gas escape prior to the shutdown of the valve #2. We thus confirmed that our modifications greatly reduced the escape of impact-generated gases from the experimental chamber through valve #2 at similar impact velocities.

The impact velocities and target porosities for all shots are summarised in Table S1. In shot numbers #447, #448, #459, and #461, we inserted a cold trap unit with liquid nitrogen into an experimental chamber to reduce the background level of water vapour. The estimated corresponding impact velocities in the main asteroid belt region calculated in Supplementary Note S7 are also included in the table.

**Supplementary Table S1.** Experimental conditions. Shot numbers #440, #447, and #448, and #459 and #461, are classified as "high $v_{imp}$" and "low $v_{imp}$" in the main text.

| Shot number | $v_{imp}$ (km s$^{-1}$) | Corresponding $v_{imp}$ (km s$^{-1}$) [a] | Target density (Mg m$^{-3}$) | Porosity (%) |
|---|---|---|---|---|
| #440 | 5.90 | 6.0–7.0 | 1.91 | 24 |
| #447 | 5.77 | 6.0–7.0 | 1.95 | 22 |
| #448 | 5.75 | 6.0–7.0 | 1.91 | 24 |
| #459 | 3.64 | 4.0–4.5 | 1.90 | 24 |
| #461 | 3.77 | 4.0–4.5 | 1.92 | 23 |

a. See main text.

**Supplementary Note S2. Sample description**

 We used a carbonaceous chondrite analogue (CI simulant) with a mineral composition similar to the Orgueil meteorite made by Exolith (https://sciences.ucf.edu/class/exolithlab/), as mentioned in the main text. Although the constituents of the simulant are summarised in Britt et al. (2019)[3], we briefly describe the simulant in this section. The mineral composition is listed in Supplementary Table S2. The main constituent is Mg-serpentine with a mass fraction of 48 wt.%. All the carbon in the simulant is contained in sub-bituminous coal with a carbon content of ~40 wt.%. Based on the van Krevelen diagram[4], the C/O and C/H atomic ratios of this type of sub-bituminous coal are ~5 and ~1.1, respectively. Magnetite is present as iron oxides. Phyllosilicates, including palygorskite and vermiculite, are also present in the simulant.

 Britt et al. (2019)[3] conducted a thermogravimetric analysis of the simulant. They heated the simulant from room temperature (300 K) to 1300 K at 35 K min$^{-1}$ and measured the change in the sample mass and composition of the released gases with increasing temperature [fig. 6 in Britt et al., 2019[3]]. From 300 to 700 K, the released gas is water vapour, which is expected to be adsorbed and/or interlayer water. The structural water in palygorskite and vermiculite should also be released at low and moderate temperatures[5]. The mass fraction of the released water at this temperature range is ~2 wt.% of the sample mass. Massive devolatilisation of the simulant initiates around 800 K. At temperatures of >700 K, a second pulse of water vapour release was observed until ~1300 K with a peak at ~1000 K. The emission of trace amounts of hydrocarbons started at ~700 K and reached a peak at ~850 K. $CO_2$ emission started at ~800 K. $CO_2$ has emission peaks at 950 and 1000 K. The total volatile content was estimated to be 14.3 wt.%. The first- and second-most abundant species were $H_2O$ and $CO_2$, respectively. The C/O atomic ratio of the coal is much larger than unity as mentioned above. Nevertheless, the carbon in the coal was oxidised to $CO_2$, and not CO, indicating that additional oxidants are probably supplied by dehydration of Mg-serpentine and thermal breakdown of magnetite.

**Supplementary Table S2.** The mineral composition of the carbonaceous chondrite simulant. The values were taken from Britt et al. (2019).

| Mineral name | Mass fraction, wt.% |
|---|---|
| Mg-serpentine | 48.0 |
| Magnetite | 13.5 |
| Vermiculite | 9.0 |
| Olivine | 7.0 |
| Pyrite | 6.5 |
| Epsomite | 6.0 |
| Sub-bituminous coal | 5.0 |
| Palygorskite | 5.0 |

**Supplementary Note S3. Quantification of the impact-generated species**

To quantify the amount of each impact-generated species, we conducted a calibration experiment. We used a gas injection device to simulate impact gas emission. The device has two independent gas inlets, a vacuum chamber of known volume, and an automatic gate valve. The gas inlets were connected to gas cylinders of pure $CO_2$ and an Ar–$CH_4$ mixture. The volume mixing ratio of $CH_4$ to Ar in the latter cylinder is 1.01%. As such, we were able to produce a gas mixture with a known volume having a wide range of $CO_2$ to Ar–$CH_4$ ratios. A pressure gauge was used to monitor the partial pressure of each gas. We used $CO_2$ gas in the calibration because $CO_2$ was the main product in all the shots (Fig. 2 in the main text). The Ar–$CH_4$ mixture was chosen as a dilutant gas. The small amount of $CH_4$ in the dilutant gas (Ar) allowed us to assess the difference in the sensitivity of the QMS to different gas species.

The device injects a gas mixture of Ar–$CH_4$–$CO_2$ with a known volume and mixing ratio into the experimental chamber in a short time period (~0.1 s). The emitted gas mixture in the experimental chamber was introduced into the QMS by the argon gas flow at 500 Pa in the same way as in the actual experiments. The total pressure of the device chamber prior to gas emission was fixed at $10^4$ Pa. Since the injected gas volume is much smaller than the total volume of the experimental chamber, the pressure fluctuation due to the gas emission is <1%. For example, the total injected amounts of $CO_2$ and $CH_4$, where the partial pressure of $CO_2$ was 6 kPa, were 49 and 0.33 µmol, respectively. Hereafter, the ion current for the mass number $M/Z = i$ is denoted as $I_i$ as in the main text. Supplementary Figure S1 shows examples of the temporal variations of the ion current ratios of $I_{44}$ ($CO_2^+$) to $I_{40}$ ($Ar^+$) after the gas emissions with different mixing ratios of $CO_2$, along with the actual data obtained in shot #448. The time-series profiles due to the gas emissions in the calibration experiment were similar to the actual data. To quantify the amount of $CO_2$, we integrated the time-series profile of the ion current ratio of $I_{44}$ to $I_{40}$ against time. The integration was carried out from $t = 0$ s, where $t$ is the time after the impact, to the time when the current ratio decreases to 20% of the peak value. Supplementary Figure S2 shows an example of the integration of the ion current ratio obtained for shot #448. The red shaded region is defined as the integrated area. Supplementary Figure S3 shows the moles of injected $CO_2$ as a function of the integrated area, which demonstrates that the integrated area is linearly proportional to the injected amount of $CO_2$. Given that the integrated area must be zero when there is no gas injection, we fitted the 24 data points with a linear function (i.e., $y = ax$, where $y$ is the injected moles of $CO_2$, $a$ is a constant determined by the linear fitting, and $x$ is the integrated area). We obtained $a = 518 \pm 19$ ($2\sigma$) and used this value to obtain the amount of $CO_2$ released by each shot. The uncertainty is $\pm 7.5\%$. As mentioned in the main text, we assumed that the ratios of the other detected gases to $CO_2$ in the integrated area correspond to the molar ratio of each species to $CO_2$. Supplementary Figure S4 shows the relationship between the molar ratio of $CH_4$ to $CO_2$ in the injected gas mixture and the ratio for the integrated area of $CO_2$ to $CH_4$. Although the data exhibit some scatter, the correlation is approximately 1:1. Thus, the above assumption is at least valid for the molar ratio of $CH_4$ to $CO_2$. Supplementary Table S3 is a summary of the impact experiment data, which supports Fig. 2 in the main text.

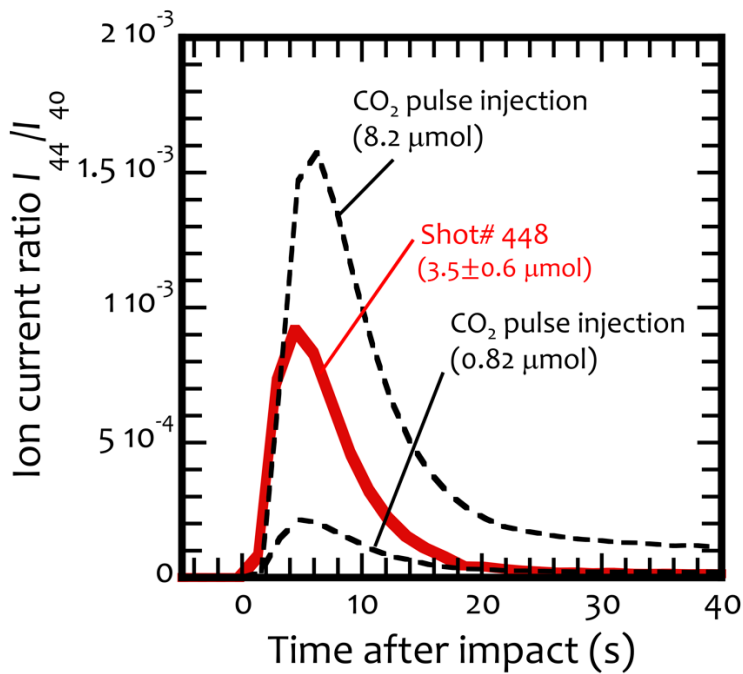

**Supplementary Figure S1.** Temporal variations of the ion current ratios $I_{44}$ to $I_{40}$ for shot #448 (i.e., high $v_{\text{imp}}$; red solid line). The results of the calibration experiments with two different amounts of injected $CO_2$ are also plotted as dashed black lines.

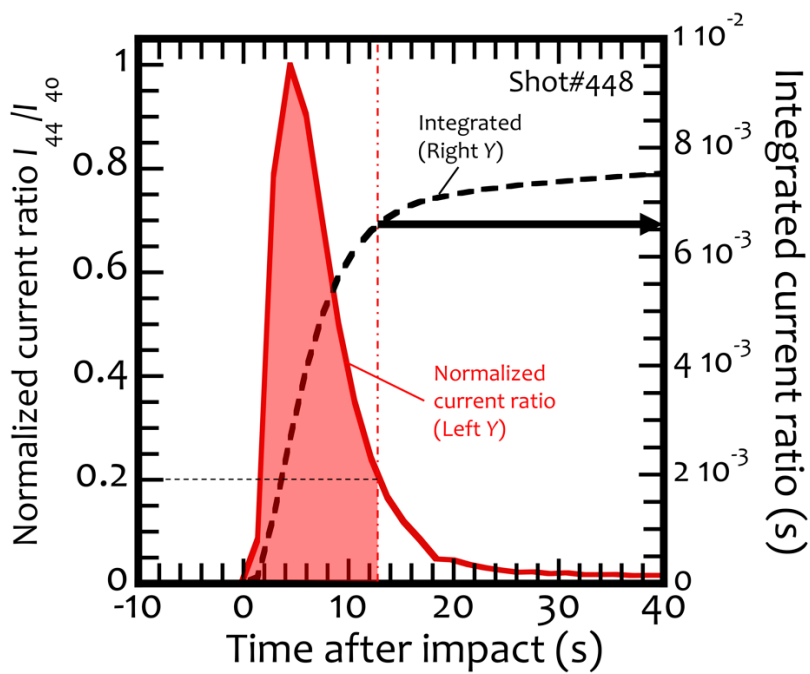

**Supplementary Figure S2.** The procedure for calculating the integrated area of the ion current ratio. In this figure, we used the results for shot #448 as an example. The normalised ion current ratio of $I_{44}$ to $I_{40}$ is shown as a function of time on the left-hand $y$-axis. The integrated area from t = 0 to a given time is plotted as a dashed black line on the right-hand $y$-axis. The red shaded region is the integrated area (see Supplementary Note S2).

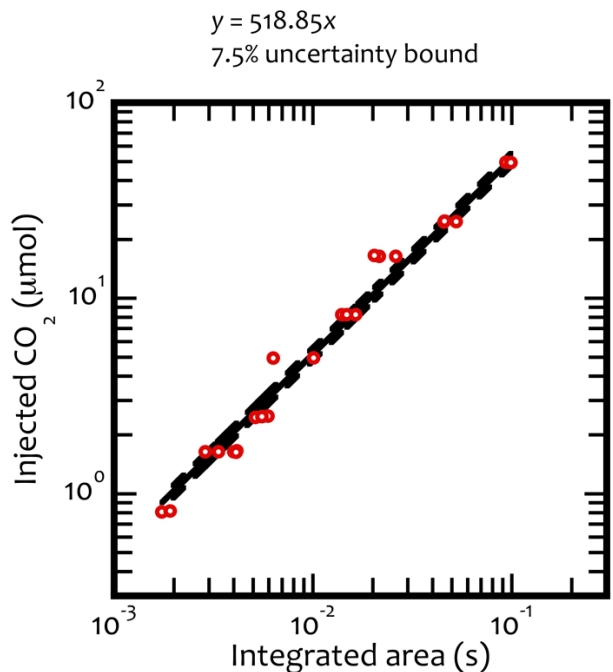

**Supplementary Figure S3.** Results of the calibration experiments using the gas injection device. The relationship between the amount of injected $CO_2$ and the integrated area is shown (open red circles). The best-fit line (see Supplementary Note S2) with a $2\sigma$ uncertainty bound is also shown as a black line.

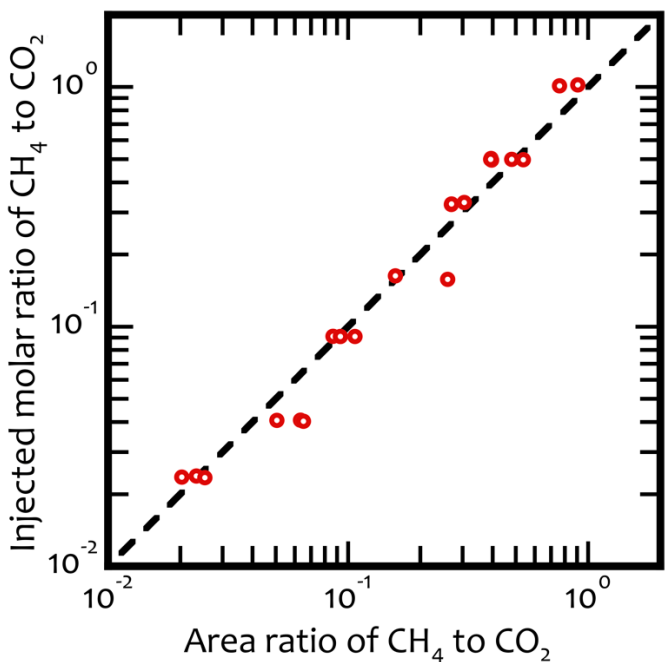

**Supplementary Figure S4.** Relationship between the injected molar ratio of $CH_4$ to $CO_2$ and the integrated area ratio of $CH_4$ to $CO_2$. The black dashed line indicates a 1:1 correlation.

**Supplementary Table S3.** Amounts of each species released by the impact experiments. The uncertainty is $\pm 7.5\%$.

| Shot number | $v_{imp}$ (km s$^{-1}$) | $H_2$ (μmol) | $CH_4$ (μmol) | CO (μmol) | $H_2S$ (μmol) | $CO_2$ (μmol) | $CO_2$ mass ($M_p^a$; %) |
|---|---|---|---|---|---|---|---|
| 440 | 5.90 | 1.0 | 0.56 | 0.77 | 0.028 | 5.8 | 1.5 |
| 447 | 5.77 | 1.8 | 0.49 | 1.5 | 0.016 | 3.6 | 0.95 |
| 448 | 5.75 | 1.1 | 0.49 | 1.0 | 0.0044 | 3.5 | 0.93 |
| 459 | 3.64 | 0.19 | 0.12 | 0.16 | N.D.[b] | 0.98 | 0.26 |
| 461 | 3.77 | 0.29 | 0.072 | 0.35 | N.D.[b] | 0.67 | 0.18 |

a. Projectile mass = 16.5 mg.
b. Not detected.

As mentioned in the main text, we did not detect a rise in $I_{18}$ even at high $v_{imp}$. Although Kurosawa et al. (2019)[2] detected impact-induced water loss of gypsum ($CaSO_4 \cdot (2H_2O)$), they reported that the temporal variation of $I_{18}$ was different from the other gases. In particular, the required time for reaching the peak of $I_{18}$ is much longer than for the other gases. Thus, we also conducted another calibration experiment with a gas mixture of $H_2O$ and $CO_2$ to constrain the detection limit for water vapour in the experimental system. A small metal container filled with distilled water was connected to the gas injection device. The container was heated to ~350 K. Water vapour was introduced into the gas injection device up to the vapour pressure. The $CO_2$ gas was also used to pressurise the device. Supplementary Figure S5 shows the results of the injection of a $H_2O$–$CO_2$ mixture. We found that the detection limit of water vapour under our experimental condition was ~60 μmol. This detection limit is 1–2 orders of magnitude more than the total gas production at high $v_{imp}$ (~5 μmol). Consequently, the mass of released water vapour in the impact experiments did not exceed 7 wt.% of the projectile mass, even under the high $v_{imp}$ condition.

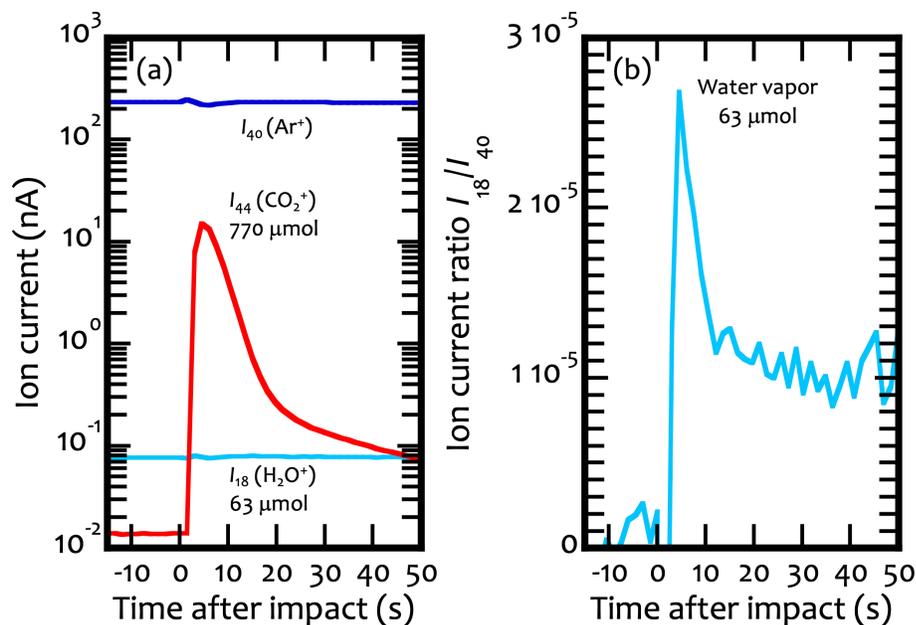

**Supplementary Figure S5**. (a) The temporal variations of the ion currents for $I_{18}(H_2O^+)$, $I_{40}(Ar^+)$, and $I_{44}(CO_2^+)$. (B) The normalised ion current ratio of $I_{18}$ to $I_{40}$.

### Supplementary Note S4. Data reduction

Our QMS uses electron impact as the ionisation method (typically referred to as the EI method). Electron bombardment of the injected gas in the QMS causes chemical cracking. For example, a certain fraction of $CO_2$ is decomposed into CO and O during the QMS measurement. Thus, $I_{16}$ and $I_{28}$ are contaminated by the cracked $CO_2$. In the same manner, $H_2O$, which is the main resident gas in the experimental chamber, is cracked into $H_2$, O, and OH. Such contamination needs to be deducted from the raw ion currents to obtain the true amounts of detected species, which are $H_2$ ($M/Z = 2$), $CH_4$ ($M/Z = 16$), and CO ($M/Z = 28$). Given that contamination of $H_2S$ and $CO_2$ due to the cracking of heavier species is unlikely, we now describe how we removed the contributions from the cracking of heavier species to the ion currents $I_2$, $I_{16}$, and $I_{28}$. We summarise the relationship between the gas species and ion currents $I_i$, including the cracking contributions to each ion current in Supplementary Figure S6.

Firstly, the ion current ratios of the cracked fractions, hereafter referred to as the cracking pattern, were measured with the QMS. We measured $CH_4$, $H_2O$, CO, $C_2H_2$, $C_2H_4$, $C_2H_6$, and $CO_2$ to obtain the cracking patterns of each standard gas. In addition, we also measured the ion current ratios of the cracked fractions of air to assess the degree of contamination from residual air in the experimental chamber. Secondly, we analysed the temporal variations of the ion currents obtained for the actual shots based on the cracking patterns. We used the average values of $I_i$ prior to the impacts (from –30 to –5 s) as background values, and these were subtracted from the raw ion currents. Supplementary Figure S7 shows an example of the data reduction pertaining to $I_{28}$. We considered contamination from impact-generated $CO_2$ and $C_2$ hydrocarbons ($C_2H_2$, $C_2H_4$, and $C_2H_6$) and molecular nitrogen ($N_2$) from the residual air in this analysis. The contamination from $CO_2$ on $I_{28}$ can be calculated from the $I_{44}$ and the cracking pattern measured using the standard $CO_2$ gas. The contamination from $C_2H_4$ on $I_{28}$ was inferred from $I_{26}(C_2H_2^+)$ and the cracking pattern of the standard $C_2H_4$ gas. The contributions from other $C_2$ hydrocarbons on $I_{28}$ are minor (not shown in the figure). The contamination from residual air ($N_2$ and $O_2$) in the experimental chamber on $I_{28}$ was inferred from $I_{32}(O_2^+)$. As shown in Fig. 1 in the main text, $I_{32}$ was almost constant during the measurements. Given that $I_{32}$ comes only from the residual air, an upper bound of the contribution of residual air to $I_{28}$ can be estimated based on the ion current ratio $I_{28}$ to $I_{32}$ obtained by measuring the air. We found that the contributions from $C_2$ hydrocarbons and the residual air to $I_{28}$ are negligible, and $I_{28}$ from the cracked $CO_2$ is also not responsible for the observed $I_{28}$. Consequently, the excess $I_{28}$ from the cracked $CO_2$ is expected to originate from the impact-generated CO vapour. We integrated the excess $I_{28}$ profile to obtain the total production of CO.

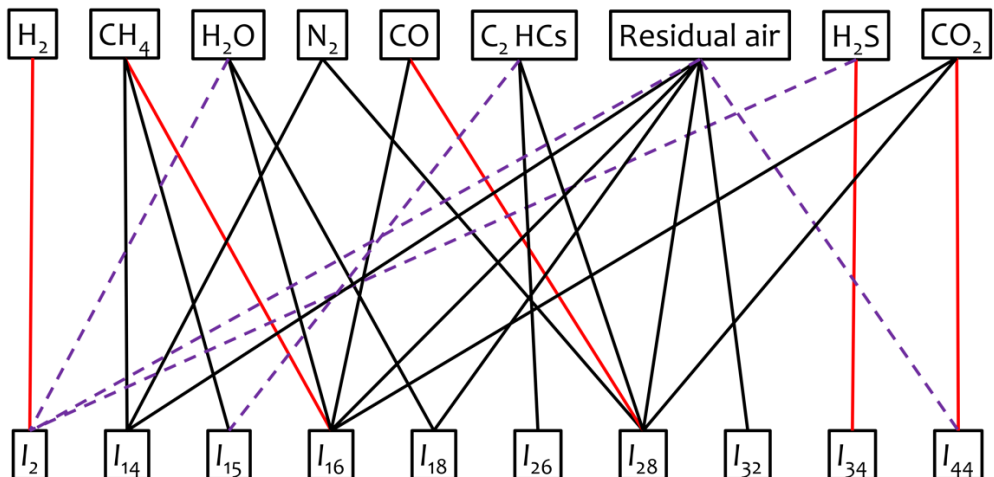

**Supplementary Figure S6.** Relationships between the gas species and ion currents. For example, $CO_2$ gas ($M/Z = 44$) yields not only $I_{44}$, but also $I_{16}$ and $I_{28}$ owing to the fragmentation by electron impacts in the QMS (see the main text). Solid red lines indicate the relationship between the detected gas species in the impact experiments and the ion currents with corresponding mass numbers. Dashed purple lines are negligible in the data reduction (see Supplementary Note S4). Note that $C_2$ HCs refers to $C_2$ hydrocarbons, which are $C_2H_2$, $C_2H_4$, and $C_2H_6$.

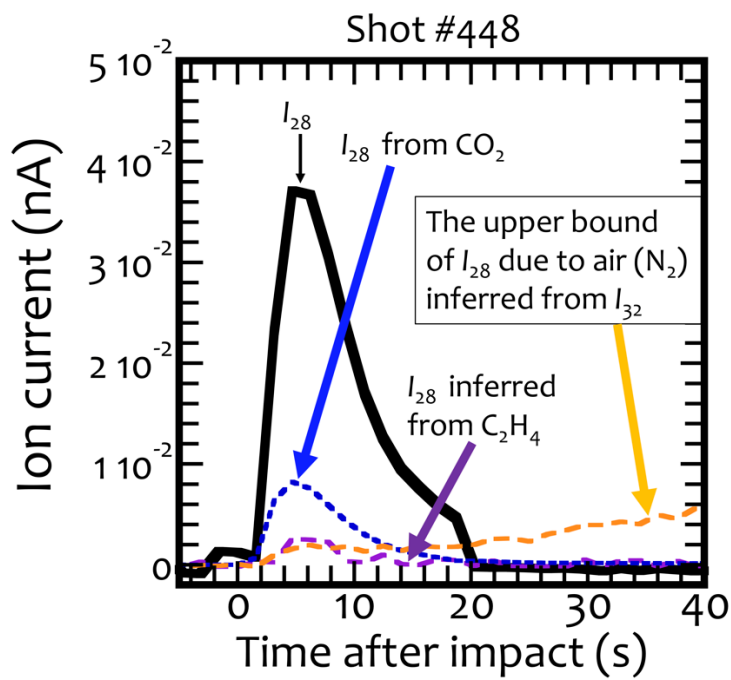

**Supplementary Figure S7.** Data reduction for $I_{28}$ for shot #448. The solid black line is $I_{28}$ after background subtraction. The contributions from the gases, except for CO, to $I_{28}$ are also shown as dashed lines (see Supplementary Note S4).

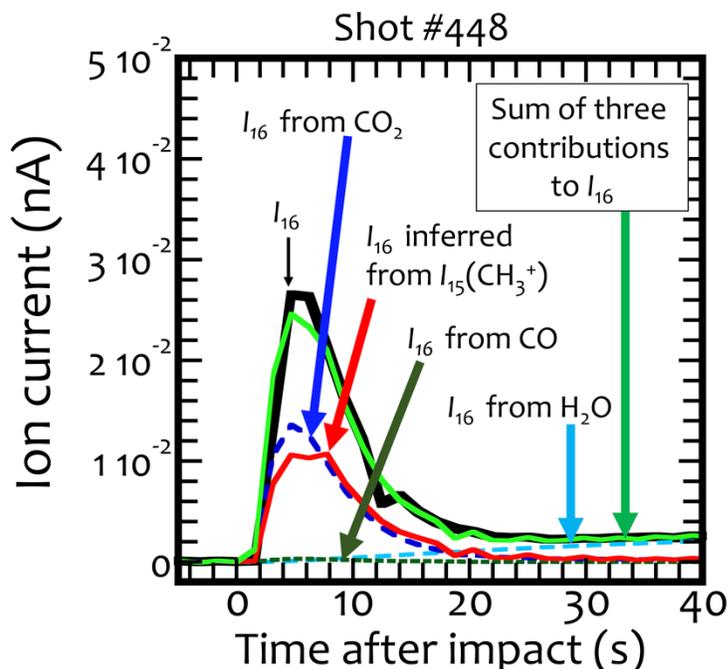

**Supplementary Figure S8.** Same as for Supplementary Fig. S7, except the results for $I_{16}$ are shown.

Supplementary Figure S8 is the same as Supplementary Fig. S7, except the data reduction pertaining to $I_{16}$ is shown. The basic data reduction concept is the same as that used for $I_{28}$. Although the contributions from the cracked $H_2O$, CO, and $CO_2$ to $I_{16}$ were considered, the contributions from $H_2O$ and CO are negligible. Approximately 50% of the $I_{16}$ comes from the cracked $CO_2$. Based on the measurement of the $CH_4$ standard, $I_{15}(CH_3^+)$ is a better indicator of the total production of $CH_4$ than $I_{16}$, because a significant fraction of $CH_4$ is cracked into $CH_3^+$ by the QMS and contamination from heavier species to $I_{15}$, such as $C_2$ hydrocarbons, is negligible. The contribution of the impact-generated $CH_4$ to $I_{16}$ was inferred from $I_{15}$, as shown in Supplementary Fig. S8, based on the cracking pattern of the $CH_4$ standard. The summation of all the contributions reproduces well the $I_{16}$ profile, suggesting that the excess $I_{16}$ from cracked $CO_2$ is due to impact-generated $CH_4$. The total production of $CH_4$ was calculated by the integration of the $I_{16}$ profile inferred from $I_{15}$ (Supplementary Fig. S8).

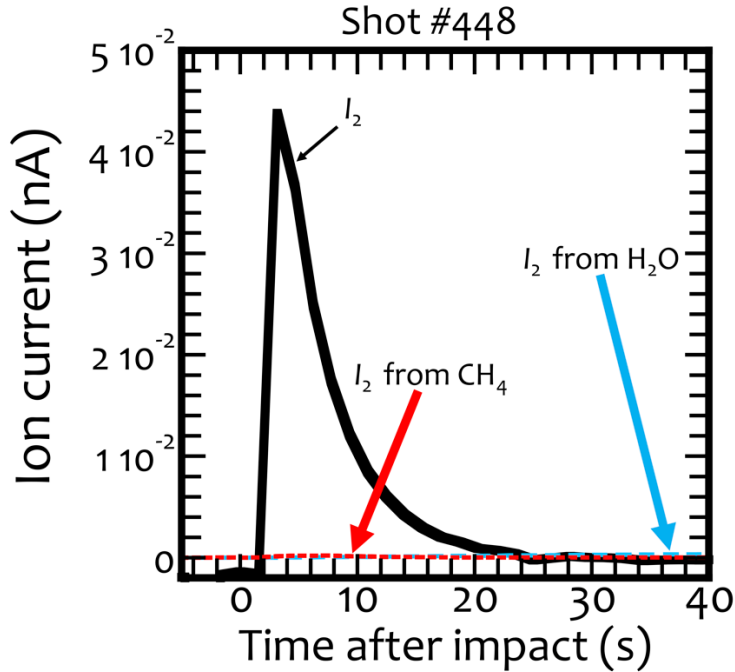

**Supplementary Figure S9.** Same as for Supplementary Fig. S7, except the results for $I_2$ are shown.

Supplementary Figure S9 is the same as Supplementary Fig. S7, except the data reduction pertaining to $I_2$ is shown. We considered the contamination from $CH_4$ and $H_2O$ to $I_2$. However, this contamination is negligible (Supplementary Fig. S9). Thus, the sudden rise in $I_2$ at around t = 0 reflects the generation of molecular hydrogen ($H_2$) due to the impact. The total production of $H_2$ was obtained by calculating the integrated area of $I_2$ in the same manner as for the other species.

**Supplementary Note S5. Comparison with the control experiment**

To verify the detected species were actually produced by hypervelocity impacts onto the target, we also conducted a control experiment (shot #442) using a basalt block (200 × 200 × 100 mm), which is not expected to contain any volatiles. The other impact conditions were mostly identical to the shots classified into the high $v_{imp}$ type. The actual impact velocity of shot #442 was 5.74 km s$^{-1}$, which is almost identical to shot #448 (5.75 km s$^{-1}$). Thus, the injected kinetic energy into the experimental chamber for the control experiment is essentially the same as for shot #448. Supplementary Figure S10 shows a comparison of the ion current ratios of selected species to $I_{40}$ for shots #448 (the CI meteorite-like material) and #442 (the basalt block). Ion current ratios are shown here because the background level of $I_{40}$ slightly fluctuated in the two different shots. The ratio $I_{15}/I_{40}$ is shown because $I_{15}$ is a better indicator of $CH_4$ than $I_{16}$, as described in Text S4. Although small rises in $I_2$, $I_{15}$, and $I_{44}$ were observed in the control experiment, the peak values were much smaller than the actual shot. In contrast, the peak value of $I_{28}$ in the control experiment is similar to that for shot #448. We conducted an in-depth analysis of the $I_{28}$ obtained in the control experiment. Supplementary Figure S11 is the same as Supplementary Fig. S7, except the results for shot #442 (control) are shown. We observed a sudden rise in $I_{14}$ for this shot, whereas hydrocarbons were not detected. This $I_{14}$ rise can be explained by the cracking of $N_2$. The $I_{28}$ due to $N_2$ was calculated from the $I_{14}$ profile and cracking pattern of air. The $I_{28}$ profile inferred from the $I_{14}$ matches the actual $I_{28}$ (Supplementary Fig. S10). Consequently, an emission of $N_2$ occurred immediately after the impact in the experimental chamber for this shot. Although the reason as to why we detected $N_2$ from the basalt block remains unclear, we can conclude that the detected species, including $H_2$, $CH_4$, CO, $H_2S$, and $CO_2$, were actually produced from the CI meteorite-like material.

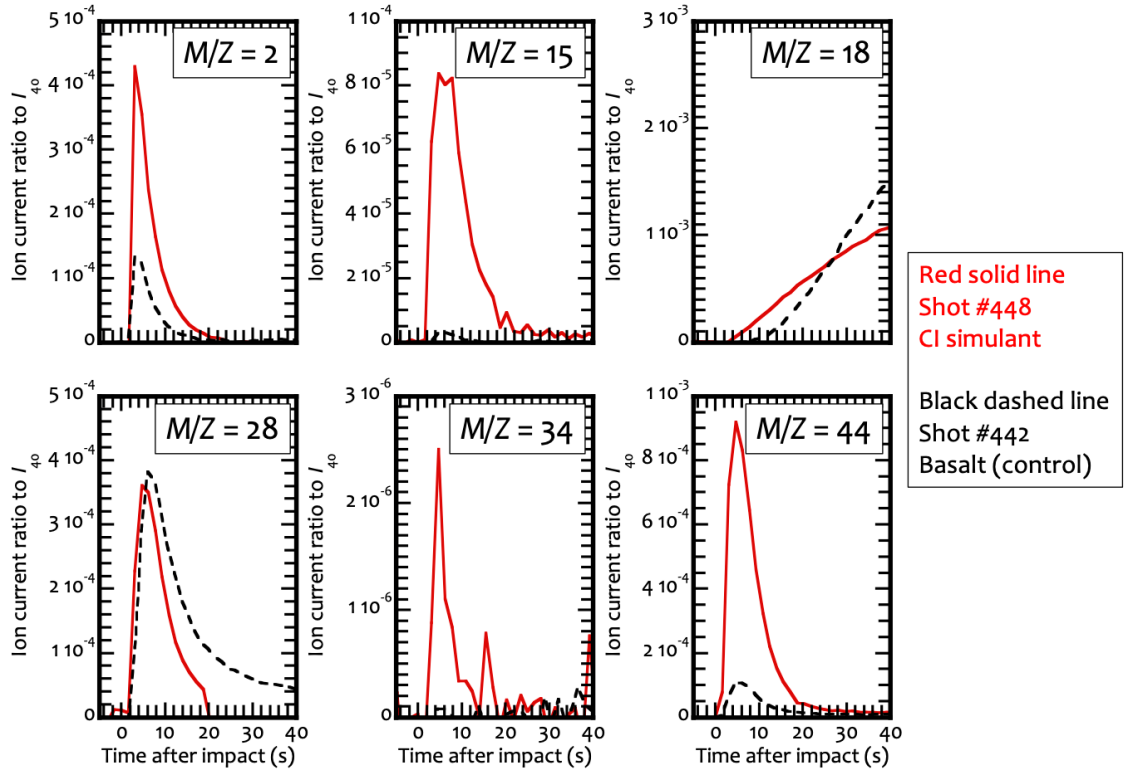

**Supplementary Figure S10.** Comparison of shot #448 with a control experiment. The results using the CI meteorite-like material and a basalt block (control) are plotted as the solid red and dashed black lines, respectively. The ion current ratios of the given mass numbers to $I_{40}$ ($Ar^+$) are plotted. The mass number for each panel is shown on the figure. Note that the upper limits of the $y$-axes for each panel are different.

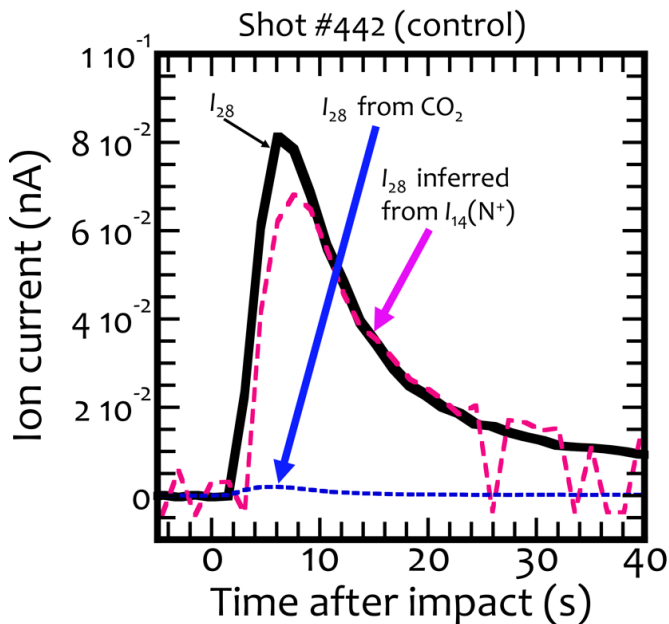

**Supplementary Figure S11.** Same as for Supplementary Fig. S7, except the results for shot #442 (i.e., the control) are shown.

**Supplementary Note S6. Calculation settings and input parameters in the iSALE computation**

We conducted two types of numerical simulations using the iSALE shock physics code[6-8]. One was carried out to estimate pressures and temperatures in the impact experiment (hereafter referred to as the Lab–simulation). In the second, we considered a collision between a planetesimal with a diameter of 20 km with the hypothesised Ryugu parent body with a 100 km diameter (hereafter referred to as the Ryugu–simulation).

In the Lab–simulation, for simplicity, we assumed that the $Al_2O_3$ projectile has no strength. A target consisting of the CI meteorite-like material was approximated using the EOS model for serpentine, which is the major constituent of this target (~50 wt.%[3]). We introduced a microscopic porosity into the target, in order to match the bulk densities of the targets used in the laboratory experiment. Given that the CI meteorite-like material (i.e., pellets) used in the laboratory experiment were made from a powder, which is a mixture of a variety of fine-grained minerals and coal, we used a simple strength model for granular materials[9-10] to account for dry friction and plastic deformation in the target. Shear deformation acting against the strength also contributes to impact heating[11]. This strength effect was also included in the computation. The input parameters for the iSALE calculations are summarised in Supplementary Table S4.

We used the same model parameters listed in Supplementary Table S4 for both the projectile and target in the Ryugu–simulation. The specific impact energy $Q$ was varied from 1 to 400 kJ kg$^{-1}$, which is close to the range of the required energy densities between cratering and catastrophic disruption $Q^*$ of 100-km-sized bodies[12-13]. We estimated the corresponding impact velocities of $v_{imp}$ achieved in the impact experiments by comparing the mass–pressure–temperature relationships obtained by two types of simulations (see Text S7 for details). The calculation settings are summarised in Supplementary Tables S5 and S6. Detailed descriptions of the variables listed in Supplementary Tables S4–S6 can be found in the iSALE manual[14].

**Supplementary Table S4.** Model input parameters for the target in the iSALE computations.

| | |
|---|---|
| EOS model | ANEOS serpentine [a] |
| Strength model | Drucker–Prager[b] |
| Poisson's ratio | 0.23 [c] |
| Melting temperature (K) | 1098 [c] |
| Thermal softening coefficient | 1.1 [d] |
| Simon parameter A (GPa) | 6 [d] |
| Simon parameter C | 3 [d] |
| Cohesion (MPa), $Y_{coh}$ | 0.01 [d] |
| Internal friction, $\mu_{dam}$ | 0.6 [d] |
| Limiting strength (GPa), $Y_{limit}$ | 1.0 [d] |
| Initial distension | 1.31 [e] |
| Elastic threshold | $-10^{-5}$ [d] |
| Transition strain | 1.0 [d] |
| Rate of porous compaction | 0.98 [d] |

a. Brookshaw (1998)[15].
b. Drucker and Prager (1952)[9]
c. Davison et al. (2016)[16].
d. Typical values were used[7].
e. Corresponding to a porosity of 25%.

**Supplementary Table S5.** Calculation settings in the iSALE calculations used for modelling the impact experiment (Lab–simulation).

| | |
|---|---|
| Computational geometry | Cylindrical coordinates |
| Number of computational cells in the $R$ direction | 1300 |
| Number of computational cells in the $Z$ direction | 1340 |
| Cells per projectile radius (CPPR) | 50 |
| Cells per target radius | 750 |
| Cells per target height | 400 |
| Artificial viscosity, $a_1$ | 0.24 |
| Artificial viscosity, $a_2$ | 1.2 |
| Impact velocity (km s$^{-1}$) | 3.7 and 5.8 |
| High-speed cutoff | Two-fold impact velocity |
| Low-density cutoff (kg m$^{-3}$) | 100 |

**Supplementary Table S6.** Calculation settings in the iSALE calculations for modelling a hypothesised impact onto the Ryugu parent body (Ryugu–simulation).

| | |
|---|---|
| Computational geometry | Cylindrical coordinates |
| Number of computational cells in the $R$ direction | 500 |
| Number of computational cells in the $Z$ direction | 1000 |
| Cells per projectile radius (CPPR) | 50 |
| Cells per target radius | 250 |
| Cells per target height | 250 |
| Artificial viscosity, $a_1$ | 0.24 |
| Artificial viscosity, $a_2$ | 1.2 |
| Impact velocity (km s$^{-1}$) | 0.5–10 at 0.5 km s$^{-1}$ steps |
| High-speed cutoff | Two-fold impact velocity |
| Low-density cutoff (kg m$^{-3}$) | 100 |

**Supplementary Note S7. Corresponding impact velocities in the main asteroid belt region**

In this study, we used an Al$_2$O$_3$ projectile with a density of 3.97 Mg m$^{-3}$, which is much denser than the CI meteorite-like material (1.9 Mg m$^{-3}$) and expected densities of C-type asteroids (<2 Mg m$^{-3}$). Thus, collisions in the main asteroid belt region at somewhat higher impact velocities were simulated to assess the temperatures that are attained. In this section, we describe the details of the estimation of the corresponding impact velocities by comparing the mass distributions (Fig. 4 in the main text) obtained from the two different simulations outlined in Supplementary Note S6.

The calculation procedure of the cumulative mass at a given specific entropy is as follows. Given that the high-temperature layer near the crater wall is relatively thin (<$D_p$) in the grid-based code, the heated materials tend to experience the effect of numerical diffusion, resulting in an artificial decrease in entropy with time. To minimise the effect of such a numerical artefact, we used the entropy value at 1 GPa stored on each tracer during decompression to yield the entropy–cumulative mass relationship. To exclude weakly shocked materials (peak pressure < 3 GPa), we only used tracers that were once shocked (>3 GPa), but were released down to 1 GPa. In other words, we did not continue the numerical integration through to complete decompression. This approach largely prevents the accumulation of numerical error[15]. Figure 4 in the main text shows clearly that our impact experiments reproduce a similar entropy–cumulative mass relationship in the case of a hypothesised catastrophic disruption of the Ryugu parent body. The corresponding impact velocities of low and high $v_{imp}$, which were achieved in the impact experiments, were estimated to be 4.0–4.5 and 6.0–7.0 km s$^{-1}$, respectively, as mentioned in the main text.

**Supplementary Note S8. The mechanism of devolatilisation**

In this section, we discuss the mechanism of impact devolatilisation of the CI simulant based on the experimental and simulation results.

*S8.1. Thermodynamics of high-temperature decomposition of the CI simulant*

We now describe the procedure for calculating the residual temperature $T_{res}$ from the entropy change that can be obtained by iSALE calculations. The current version of ANEOS serpentine used in the iSALE calculations does not include the effect of endothermic decomposition of hydrous minerals and organics on the residual temperature. Thermogravimetric analysis of the CI simulant[3] showed that dehydration of the structural water in Mg-serpentine and thermal decomposition of coal initiates at ~800 K, as mentioned in Supplementary Note S2. In the following, we assumed that the dehydration of Mg-serpentine and decomposition of coal proceeds simultaneously at a single decomposition temperature $T_d = 800$ K.

If we neglect the endothermic decomposition, the residual temperature $T_{res,sh}$, which is the temperature after the shocked material releases down to the reference pressure, is given by[16]:

$$T_{res,sh} = T_0 \exp\left(\frac{\Delta S}{C_p}\right) \text{ (S1),}$$

where $\Delta S$ is the entropy increase from the reference state and $T_0$ is the initial temperature. The entropy increase, $\Delta S$, is one of the ANEOS outputs during the iSALE calculations. We then estimated an effective value of $C_p$ at temperatures between 200–2000 K, as follows. Although the ANEOS code calculates the temperature dependence of $C_p$ with the Debye–Grüneisen model[17], the temperature dependence is rather small at sufficiently high temperatures (>>300 K). By comparing the calculated values of Eq. (S1) having a single $C_p$ value with the temperature–entropy relationship at the reference volume calculated by the ANEOS, we confirmed that a constant value of $C_p = 1.53$ kJ K$^{-1}$ kg$^{-1}$ almost reproduces the ANEOS outputs with the more accurate $C_P$ treatment. Since dehydration of hydrous minerals and decomposition of coal at $T_d$ are endothermic reactions, Eq. (S1) cannot be applied directly to the case where $T_{res,sh}$ exceeds $T_d$. In the following, we quantitatively evaluated the endothermic heat by assuming typical reactions with the constant $C_p$.

Residual enthalpy $H_{res}$, which is the heat implanted owing to irreversible heating, is estimated to be:

$H_{res} = C_p T_{res,sh}$. (S3)

Note that Eq. (S3) is valid even at $T_{res,sh} \geq 800$ K. We could quantitatively estimate $H_{res}$ by a post-analysis of the iSALE results. The actual residual temperature $T_{res}$ at $T_{res,sh} \geq T_d$, including the effect of endothermic heat, is given by:

$$T_{res} = \frac{H_{res} - H_{serp} - H_{org}}{C_p}, \text{ (S4)}$$

where $H_{serp}$ and $H_{org}$ are the dehydration and decomposition enthalpies of Mg-serpentine and the coal in the CI simulant, respectively. $T_{res,sh}$ is higher than $T_{res}$ at $\geq T_d$. If we assume the following reactions pertaining to dehydration of Mg-serpentine and decomposition of the coal

Mg-serpentine = Mg-silicates + 2H$_2$O (S5)

and

C + H$_2$O = CO + H$_2$, (S6)

respectively, $H_{serp}$ and $H_{org}$ can be quantitatively estimated. Note that the latent heat of vaporisation of water should be added to $H_{serp}$. The Mg-serpentine and coal in the heated region having $T_{res,sh} \geq T_d$ are completely dehydrated/decomposed under the condition where $T_{res}$ defined by Eq. (S4) is higher than $T_d$. The entropies required for incipient and complete devolatilisation of the CI simulant $S_{id}$ and $S_{cd}$ are estimated by

$$S_{id} = S_0 + C_p \ln\left(\frac{T_d}{T_o}\right) \quad (S7)$$

and

$$S_{cd} = S_0 + C_p \ln\left(\frac{C_p T_d - H_{serp} - H_{org}}{C_p T_o}\right), \quad (S8)$$

respectively, where $S_0$ is the reference entropy. In the case of the reactions in Eqs (S5–S6), $S_{id}$ and $S_{cd}$ are estimated to be 3.1 and 3.7 kJ K$^{-1}$ kg$^{-1}$, respectively. Supplementary Figure S12 shows the residual temperature as a function of specific entropy. In the entropy range of $S_{id} < S < S_{cd}$, the temperature field after the impact is thermally buffered by the endothermic reactions, resulting in $T_{res} = T_d = 800$ K. Figures 3 and 4b in the main text show the residual temperatures calculated from the red curve shown in Fig. S12. Figure 4 in the main text and Supplementary Fig. S12 show that complete devolatilisation of the CI simulant is unlikely even at the impactor footprint at low $v_{imp}$, which is the condition that corresponds to typical collisions in the main asteroid belt (Supplementary Note S7). The "heat budget" of volatile-bearing constituents in the CI-chondrite-like material are sufficiently large to suppress the degree of impact devolatilisation. All the parameters used to obtain $S_{id}$ and $S_{cd}$ are listed in Supplementary Table S7. It should be noted that the decomposition reactions of insoluble organic matter (IOM) in actual carbonaceous meteorites is not expected to be as simple as assumed by Eq. (S6). The values shown here are first-order estimates.

**Supplementary Table S7.** Thermodynamic constants used to estimate the entropies required for incipient and complete devolatilisation of the CI simulant.

| | |
|---|---|
| Isobaric specific heat of CI simulant $C_p$ (kJ K$^{-1}$ kg$^{-1}$) | 1.53[a] |
| Dehydration enthalpy of serpentine (MJ kg$^{-1}$) | 0.42[b] |
| Latent heat of vaporisation of water (MJ kg$^{-1}$) | 2.26[c] |
| Dehydration enthalpy of CI simulant (MJ kg$^{-1}$) | 0.34 |
| Reaction enthalpy of Eq. (S6) (MJ kg$^{-1}$ Carbon) | 11.3[d] |
| Decomposition enthalpy of organics in CI simulant (MJ kg$^{-1}$) | 0.26 |
| Reference entropy $S_0$ at 220 K (kJ K$^{-1}$ kg$^{-1}$) | 1.15[e] |
| Entropy for incipient decomposition $S_{id}$ (kJ K$^{-1}$ kg$^{-1}$) | 3.1 |
| Entropy for complete decomposition $S_{cd}$ (kJ K$^{-1}$ kg$^{-1}$) | 3.7 |

a. Fitted to the ANEOS output values (Supplementary Note S7.1)
b. The dehydration enthalpy of Mg-serpentine was taken from Waber and Greer (1965)[18]. As mentioned in Text S8.1, the latent heat of vaporisation of water was added.
c. Haynes (2010)[19].
d. Chase (1998)[20].
e. The output of the ANEOS for serpentine[21]

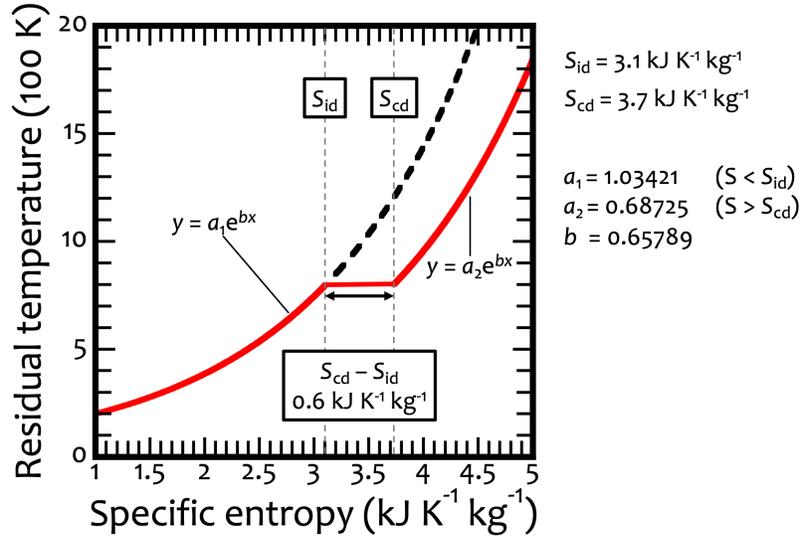

**Supplementary Figure S12.** Residual temperature as a function of entropy. Entropies required for incipient ($S_{id}$) and complete ($S_{cd}$) devolatilisation are shown as vertical dashed lines. Best-fit equations and constants are shown on the figure.

*S8.2. Constraints from the simulant composition, measured gas composition, and iSALE results*

In this section, we summarise the constraints available to explore the mechanism of impact devolatilisation of the CI simulant. First, we estimated the upper limit of water and $CO_2$ productions $N_{H2O}$ and $N_{CO2}$, based on the elemental and mineral compositions of the CI simulant and the iSALE results. Impact devolatilisation only proceeds in the heated region that has $T_{res} \geq T_d$, as mentioned in section S8.1. Thus, we used the heated target masses $M_{800K}$ with $T_{res} \geq T_d$ at high and low $v_{imp}$ as the reference masses. The upper limits of $N_{H2O}$ and $N_{CO2}$ can be estimated from the composition of the target. Assuming that all the Mg-serpentine in the heated region is dehydrated and all the carbon in the coal in the same region is oxidised to $CO_2$, the upper limit of the mass of the released gas is ~0.43 $M_p$ at high $v_{imp}$ and ~0.14 $M_p$ at low $v_{imp}$. The upper limits of $N_{CO2}$ are 93 μmol for high $v_{imp}$ and 29 μmol for low $v_{imp}$. These values are ~30 times higher than the amount of $CO_2$ actually detected (Table S3). As described in Supplementary Note S3, the detection limit of water vapour in the experiments is ~60 μmol, suggesting that the detection of water vapour was possible under our experimental conditions if all Mg-serpentine in the heated region with $T_{res} \geq T_d$ was dehydrated (160 μmol at high $v_{imp}$). These data are summarised in Supplementary Table S8.

Secondly, we summarise the constraints from the experimentally measured gas composition. The impact-generated gases for all the shots are dominantly $CO_2$ (48–71 mol.%), which shows that the coal is strongly oxidised due to oxygen derived from other sources. Such sources are expected to be Mg-serpentine and magnetite. As described in the main text, the molecular composition of the impact-generated gas ($CO/CO_2$ and $H_2/CO$ ratios) was almost constant regardless of the impact velocity (Fig. 2; Supplementary Table S3). Gas compositions of $CO/CO_2 > 0.1$ and $H_2/CO \sim 1$ provide important constraints on the temperature of the location where devolatilisation occurs as discussed in the following two sections.

**Supplementary Table S8.** Estimated upper limits of water and CO$_2$ production.

|  | High $v_{imp}$ | Low $v_{imp}$ |
|---|---|---|
| Impact velocity (km s$^{-1}$) | 5.8 | 3.7 |
| Kinetic energy $E_0$ (J)[a] | 278 | 113 |
| Heated mass of the target having $T_{res} \geq T_d$ $M_{800K}(M_p)$[b] | 2.8 | 0.85 |
| Residual enthalpy $H_{res}$ ($E_0$)[c] | 0.27 | 0.19 |
| Heated serpentine mass ($\geq$800 K) (mg)[d] | 22 | 6.7 |
| Upper limit of water production (%, $M_p$)[e] | 17 | 5.3 |
| Upper limit of water production (μmol) | 158 | 49 |
| Heated coal mass ($\geq$800 K) (mg)[f] | 2.3 | 0.70 |
| Heated carbon mole number ($\geq$800 K) (μmol)[g] | 86 | 26 |
| Upper limit of released CO$_2$ mass (%, $M_p$) | 23 | 7.0 |

a. A projectile mass $M_p$ = 16.5 mg was used.
b. The results of the Lab-simulations.
c. An isobaric specific heat $C_p$ = 1.53 kJ K$^{-1}$ kg$^{-1}$ and $T_0$ = 280 K were used.
d. A mass fraction of Mg-serpentine to the CI simulant of 48 wt.% was used.
e. An idealised chemical formula for Mg-serpentine (Mg$_3$Si$_2$O$_5$(OH)$_4$) and Eq. (S5) were assumed. The molecular weight of the Mg-serpentine is 276 × 10$^{-3}$ kg mol$^{-1}$.
f. A mass fraction of the coal to the CI simulant of 5 wt.% was used.
g. The carbon content of the coal (45 wt.%) used to make the CI simulant was assumed.

*S8.3. Devolatilisation by simple shock propagation*

In this and the following section, we discuss the degassing mechanism from the CI simulant based on the constraints described in Supplementary Note S8.2. Here we show that a simple shock–release cycle cannot explain the experimental results. Recently, Kurosawa et al. (2021)[15] developed a method, which is referred to as the two-step method, to estimate the total CO$_2$ production during impact devolatilisation of calcite. Given that $S_{id}$ and $S_{cd}$ are quantitatively estimated as described in Supplementary Note S8.1, the two-step method can be applied to impact devolatilisation of the CI simulant. The degree of devolatilisation $\psi$ of carbon in the coal to CO and/or CO$_2$ is given by the lever rule as follows:

$$\psi = \begin{cases} 0, & (S < S_{id}) \\ \frac{(S-S_{id})}{(S_{cd}-S_{id})}, & (S_{id} < S < S_{cd}) \\ 1, & (S_{cd} < S) \end{cases} \quad (S9)$$

The total CO–CO$_2$ production $N_{CO-CO2}$ is the estimated with the following equation:

$$N_{CO-CO2} = \sum \left(\frac{f_c}{\mu_C} \psi_i m_i\right), \quad (S10)$$

where $f_C = 0.023$ is the mass fraction of carbon in the CI simulant (Supplementary Note S2), $\mu_C = 12 \times 10^{-3}$ kg mol$^{-1}$ is the atomic weight of carbon, and $m$ is the mass of the tracer particle in cylindrical coordinates. The subscript $i$ indicates the tracer ID. Note that we used the value of entropy $S$ after the pressure of each tracer decreased down to 1 GPa to prevent numerical artefacts, as mentioned in Supplementary Note S7. We confirmed that the estimated $N_{CO-CO2}$ becomes constant after $t > 6\, t_s$, where $t$ and $t_s$ are the time and characteristic time for projectile penetration, respectively. The characteristic time $t_s$ is defined to be $t_s = 2R_p/v_{imp}$, where $R_p$ is the projectile radius. The $N_{CO-CO2}$ value at the time when $N_{CO2}$ became constant was used as $N_{CO-CO2,calc}$ in the following. Supplementary Figure S13 shows $N_{CO-CO2}$ as a function of $v_{imp}$. The experimental $N_{CO-CO2,exp}$ and iSALE results $N_{CO-CO2,calc}$ are shown, along with the upper limit of the total production of carbon-bearing gases. The iSALE results are one order of magnitude greater than the experimental results. Nevertheless, the $v_{imp}$ dependence of $N_{CO-CO2}$ (i.e., $N_{CO-CO2} \propto v_{imp}^{3.5}$) can be reproduced by the two-step method, indicating that the degree of irreversible heating (entropy gain) is an important controlling factor on total gas production.

We now discuss the gas composition. As mentioned in the main text, CO was detected regardless of the impact velocity, and the CO/$CO_2$ molar ratios of the impact-generated gases did not depend on the impact velocity (Fig. 2; Supplementary Table S3). Given that the bulk atomic C/O ratio of the CI simulant is <1, $CO_2$ is stable at relatively low temperatures[22,23]. Therefore, the detection of CO suggests that the degassing region was strongly heated. All the carbon in the detected gases (CO and $CO_2$) must have been derived from the coal in the CI simulant because it does not contain any other carbon-bearing minerals. The CO/$CO_2$ molar ratio is simply controlled by the oxygen fugacity $f_{O2}$ of the heated region, where decomposition of the coal proceeds as follows[24]:

$$\frac{N_{CO}}{N_{CO2}} = \frac{K_{CO-CO2}}{f_{O2}^{1/2}}, \quad (S11)$$

where the left side of the equation is the molar ratio of CO/$CO_2$ and $K_{CO-CO2}$ is the equilibrium constant of the reaction $CO_2 = 2CO + O_2$. Schaefer and Fegley (2017)[24] showed that the pressure dependence of $f_{O2}$ is rather small, and described $K_{CO-CO2}$ and $f_{O2}$ as a function of temperature as follows:

$$\log_{10} K_{CO-CO2} = -\frac{14,787}{T} + 4.5472 \quad (S12)$$

and

$$\log_{10} f_{O2} = a + b\frac{10^3}{T} + c\frac{10^6}{T^2} + d\frac{10^9}{T^3} + f\frac{10^{12}}{T^4}, \quad (S13)$$

where $T$ is temperature and $a = 2.4976$, $b = -9.8605$, $c = -17.0701$, $d = 7.5220$, and $f = -1.0404$. The five constants were determined from the elemental composition of CI chondrites[24]. Supplementary Figure S14 shows the CO/$CO_2$ molar ratio as a function of temperature under the oxygen fugacity conditions of the CI composition. The measured CO/$CO_2$ molar ratios in the impact experiments are also shown as dotted horizontal lines. Figures 3 and 4 in the main text clearly show that the residual temperature field is largely buffered by the endothermic decomposition of the CI simulant. The calculated CO/$CO_2$ molar ratio at $T_d$ is 0.017 and is one order of magnitude smaller than the measured values. This is also inconsistent with the experimental constraints described in Supplementary Note S7.2. Consequently, this simple thermodynamic consideration fails to explain both the total gas production and chemical composition of the impact-generated gas.

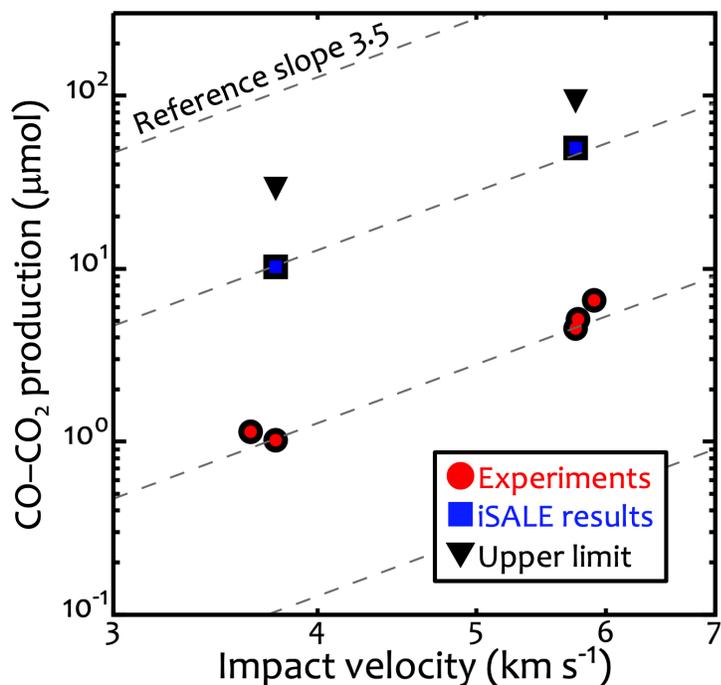

**Supplementary Figure S13.** The same as Fig. 2 in the main text, except that the sum of CO and $CO_2$ production is plotted (Supplementary Note S8.3). Note that both the *x*- and *y*-axes are shown in a log scale. Reference slopes, which represent a power law with an exponent of 3.5, are shown as dashed grey lines. The upper limits of CO and $CO_2$ production are listed in Supplementary Table S8.

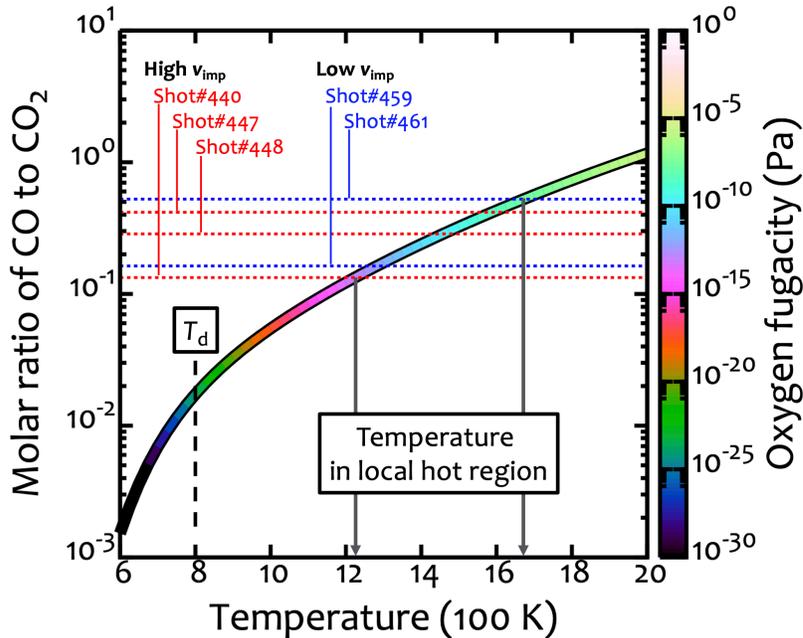

**Supplementary Figure S14.** Molar ratio of CO to $CO_2$ as a function of temperature for the elemental composition of CI chondrites. The colour bar indicates oxygen fugacity. Measured $CO/CO_2$ ratios from the impact experiments (Supplementary Table S2) are plotted as five dotted horizontal lines highlighted in red and blue for high and low $v_{imp}$, respectively.

*S8.4. Devolatilisation from a local hot region*

The measured $CO/CO_2$ ratio can be used to estimate the temperature of the region where devolatilisation occurs (Supplementary Fig. S14). By comparing the measured $CO/CO_2$ ratios with the calculated $CO/CO_2$ ratios (Supplementary Fig. S14), the temperatures of the location where decomposition proceeded are estimated to be 1200–1650 K, which are much higher than $T_d$ (= 800 K). The estimated temperatures do not depend on $v_{imp}$, which implies that devolatilisation of the CI simulant was caused by a self-limiting process. Bland et al. (2014) performed shock propagation simulations in a medium containing several different types of materials on a mesoscale, and showed that significant temperature heterogeneities existed. Such local energy concentrations would also occur under our experimental conditions, because the CI simulant contains a variety of different materials (Supplementary Table S2). Supplementary Figure S15 shows particle velocities behind a shock wave pertaining to several minerals in the CI simulant. As mentioned in Supplementary Note S6, we assumed that the hydrodynamic response to shock propagation in the CI simulant can be calculated from the properties of porous serpentine. However, Supplementary Fig. S15 shows that there is a large contrast in shock impedance amongst different minerals, resulting in shear velocities up to ~1 km s$^{-1}$ between different grains. Thus, the local heating mentioned above is probably due to friction between different grains. The degree of frictional heating is limited by thermal softening at high temperatures. Since the estimated temperature of the decomposition field (1200–1650 K) is similar to the solidus temperatures of the constituent minerals in the CI simulant, thermal softening is expected to occur efficiently. Thus, the measured $CO/CO_2$ molar ratio that shows no clear dependence on $v_{imp}$ is qualitatively consistent with the hypothesis that devolatilisation occurs due to the local energy concentration in response to frictional heating between different grains.

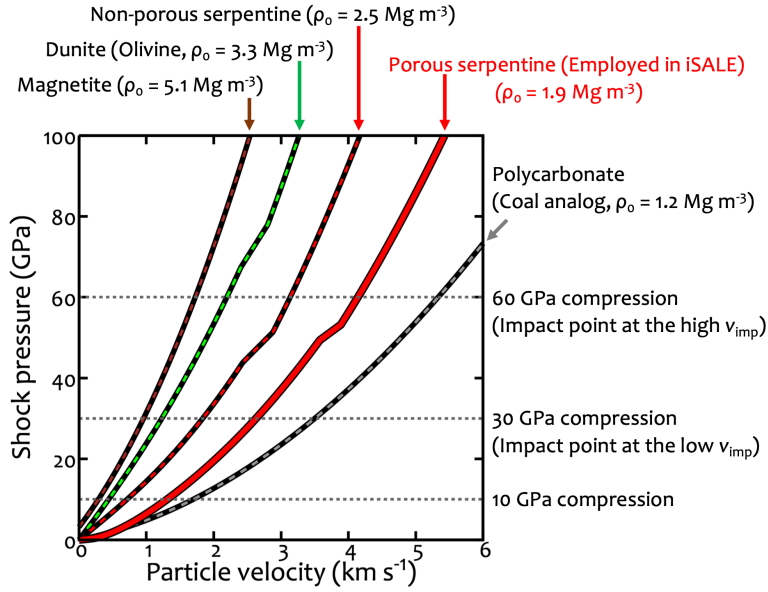

**Supplementary Figure S15**. Shock Hugoniot curves on a particle velocity–pressure plane. The curve for porous serpentine (thick red solid line) was used in the iSALE simulations. Curves pertaining to magnetite[25], dunite[26], non-porous serpentine[21], and polycarbonate[27] are also plotted. Polycarbonate was used as an analogue for coal in the CI simulant. Dotted horizontal lines correspond to the peak pressures at the impact points at high and low $v_{imp}$.

Given that all the gas detected in the impact experiment is devolatilised from the local hot region, the volume fraction of the local hot region to the region with $T_{res,sh} \geq T_d$ can be estimated by the following approximation:

$$\frac{V_{hot}}{V_{800K}} \sim \frac{N_{CO\text{-}CO2}}{N_{C,total}} = \frac{N_{CO\text{-}CO2}}{\left(\frac{f_C}{\mu_C}\right) M_{800K}}, \quad (S14)$$

where the left side of the equation is the volume fraction of the local hot region, $N_{C,total}$ is the total number of carbon atoms in the region with $T_{res,sh} \geq T_d$, and $M_{800K}$ is the mass of the region with $T_{res,sh} \geq T_d$, respectively. The values of $M_{800K}$ for high and low $v_{imp}$ were estimated with iSALE (Supplementary Table S8). The volume fraction is estimated to be 5.8 +0.12/−0.10 vol.% at high $v_{imp}$ and 3.7 ± 0.2 vol.% at low $v_{imp}$. The volume fraction increases with increasing $v_{imp}$.

The measured $H_2$/CO molar ratio varied from 0.83 to 1.30 (Supplementary Table S3), which is close to unity regardless of $v_{imp}$. The $H_2$/CO molar ratio is consistent with Eq. (S6). The detected CO and $H_2$ are likely to be the oxidation products of coal by $H_2O$ derived from Mg-serpentine. Given the $f_{O2}$ in the CI simulant and Eq. (S13), the $H_2O$ production from the local hot region can be estimated from the detected $H_2$. Based on Schaefer and Fegley (2017), as well as Eqs (S12–S13), the $H_2O$/$H_2$ molar ratio is expressed as:

$$\frac{N_{H2O}}{N_{H2}} = \frac{f_{O2}^{1/2}}{K_{H2\text{-}H2O}}, \quad (S15)$$

and

$$\log_{10} K_{\text{H2-H2O}} = -\frac{12{,}794}{T} + 2.7768, \text{ (S16)}$$

where $K_{\text{H2-H2O}}$ is the equilibrium constant of the reaction $2H_2O = 2H_2 + O_2$. The $H_2O/H_2$ molar ratio is estimated to be 7.2–11 at the temperature of the local hot region (1200–1650 K). Supplementary Figure S16 shows the measured $H_2$ production $N_{\text{H2}}$ as a function of $v_{\text{imp}}$ along with the expected $H_2O$ production $N_{\text{H2O,fO2}}$ with $N_{\text{H2}}$ and $f_{\text{O2}}$. The expected $H_2O$ production $N_{\text{H2O,fO2}}$ is smaller than the detection limit of the system. There is another expected type of $H_2O$ production $N_{\text{H2O,hot}}$, which can be obtained by assuming devolatilisation of all the structural water in Mg-serpentine in the local hot region as follows:

$$N_{\text{H2O,hot}} = 2 \frac{V_{\text{hot}}}{V_{\text{800K}}} \frac{f_{\text{serp}}}{\mu_{\text{serp}}} M_{\text{800K}}, \text{ (S17)}$$

where $f_{\text{serp}} = 0.48$, which is the mass fraction of Mg-serpentine in the CI simulant (Table S2), and $\mu_{\text{serp}} = 276 \times 10^{-3}$ kg mol$^{-1}$, which is the molecular weight of Mg-serpentine. We also plot $N_{\text{H2O,hot}}$ in Fig. S16 and found that $N_{\text{H2O,fO2}}$ is similar to $N_{\text{H2O,hot}}$.

Based on the above discussion, it is proposed that decomposition of Mg-serpentine and coal occurred only in the vicinity of the local hot region under our experimental conditions. Given that the $H_2O/H_2$ molar ratio in the local hot region was estimated to be ~10, water vapour derived from Mg-serpentine would not be the dominant source of the oxidant for the coal. A possible oxygen source is magnetite in the vicinity of the local hot region. Magnetite is decomposed into wüstite and/or metallic iron at relatively high temperatures. Thus, the released oxygen from magnetite would oxidise a gas phase. Given that the present experiment was conducted in a fully open system, the target after the impact was pulverised into fine powder and dispersed into the entire experimental chamber. Although the shocked residue of the condensed phase would be reduced due to high temperatures, it is extremely difficult to verify the reduction because of the difficulty in sample collection after the shots. Such a recovery experiment is challenging and beyond the scope of this study. Nevertheless, disappearance of the peak corresponding to magnetite in X-ray diffraction patterns has been reported for experimentally heated CM chondrites at temperatures of 1000 K[28].

Finally, we discuss the detected $CH_4$ in the experiments. According to chemical equilibrium calculations with the elemental composition of CI chondrites[22,23], $CH_4$ is only stable at relatively low temperatures. The coexistence of $CH_4$ with CO is unlikely in the same location. The thermogravimetric analysis of the CI simulant[3] showed that release of hydrocarbons from the CI simulant occurs at low temperatures (700–900 K) with respect to $CO_2$ release (800–1000 K). Consequently, the detected $CH_4$ was expected to come from outside the local hot region and be produced by pure pyrolysis of coal in the CI simulant without any oxidants.

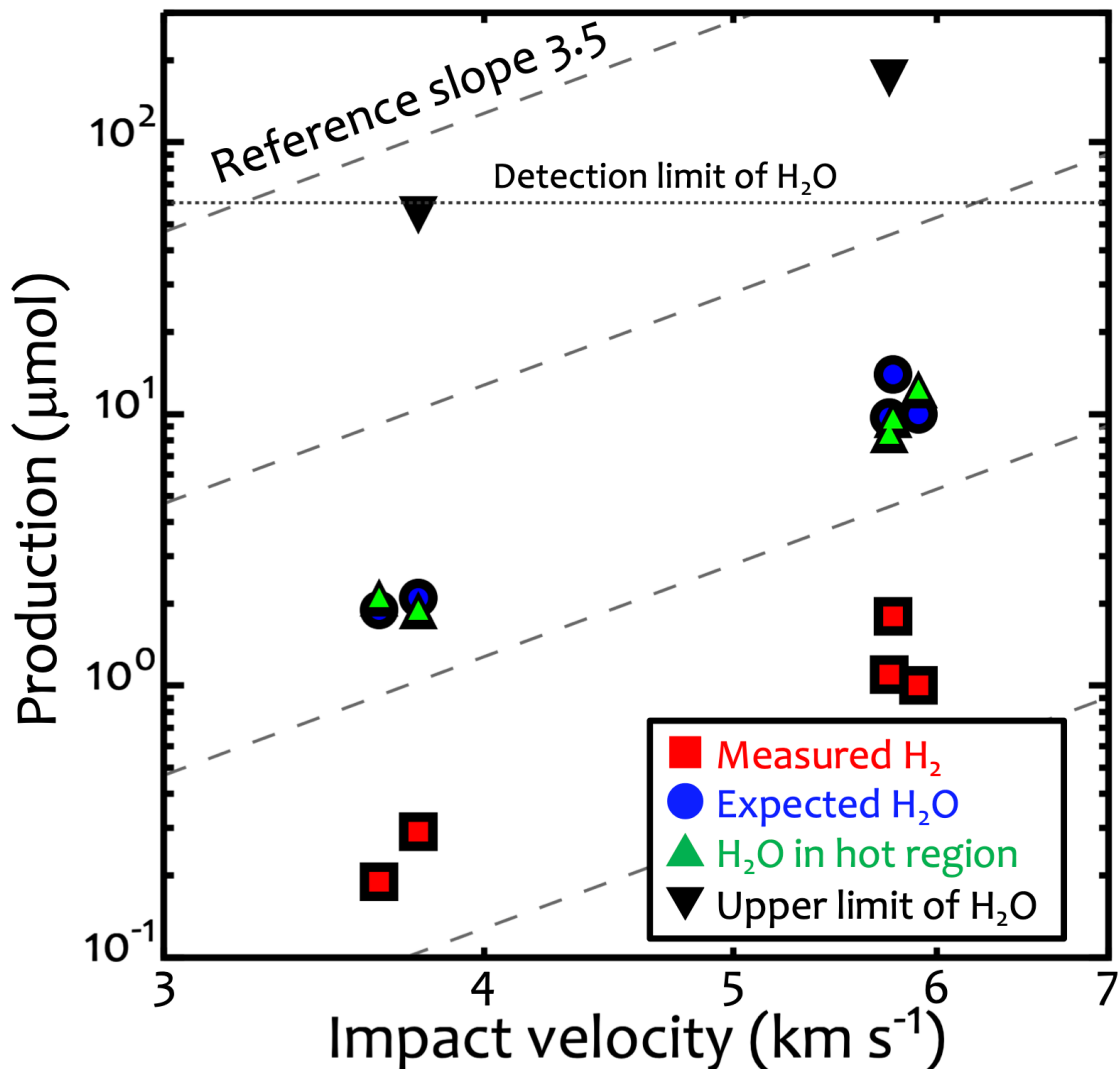

**Supplementary Figure S16.** The same as Figs 2 and S13, except that $H_2$ and $H_2O$ production are plotted (Supplementary Note S8.4). The upper limits of $H_2O$ production are listed in Table S8. The detection limit for water vapour in our system is shown as a black dotted line.

**Supplementary Note S9. Dense impactors onto a hypothesised Ryugu parent body**

As described in Supplementary Note S6, the Ryugu simulation assumes a serpentine impactor with the same porosity as the target material. However, it is also possible that the impactor of the catastrophic disruption event was an S- or M-type asteroid. In this section, we discuss the effect of impactor density on the degree of irreversible impact heating in terms of the relationship between heated mass and entropy. The calculation setting is the same as in the Ryugu-simulation, except for the EOS model of the impactor. The two calculations were performed with ANEOS dunite[26] and ANEOS iron[29]. The impact velocity was set to 5 km/s, which is the average impact velocity in the main asteroid belt region[30]. The model input parameters pertaining to rocky and metal impactors are listed in Supplementary Tables S9 and S10, respectively. Supplementary Figure S17 is the same as Fig. 4b in the main text, which is the cumulative mass at a given entropy as a function of entropy, except that the results with ANEOS dunite and ANEOS iron for impactors are also shown. It should be noted here that the cumulative mass is normalised to the impactor mass $M_p$. The mass of the iron impactor is 4.1 times that of the impactor in the Ryugu simulations. The results show that the mass–entropy distributions for the rocky and metal impactors do not significantly change from the case with an impactor of porous serpentine. In addition, our impact experiments cover the hypothesised catastrophic disruptions for possible dense impactors at the typical impact velocity in the main asteroid belt region in terms of the degree of irreversible heating. Consequently, significant volatile loss of the Ryugu parent body is unlikely to have occurred due to hypervelocity impacts in the main asteroid belt region.

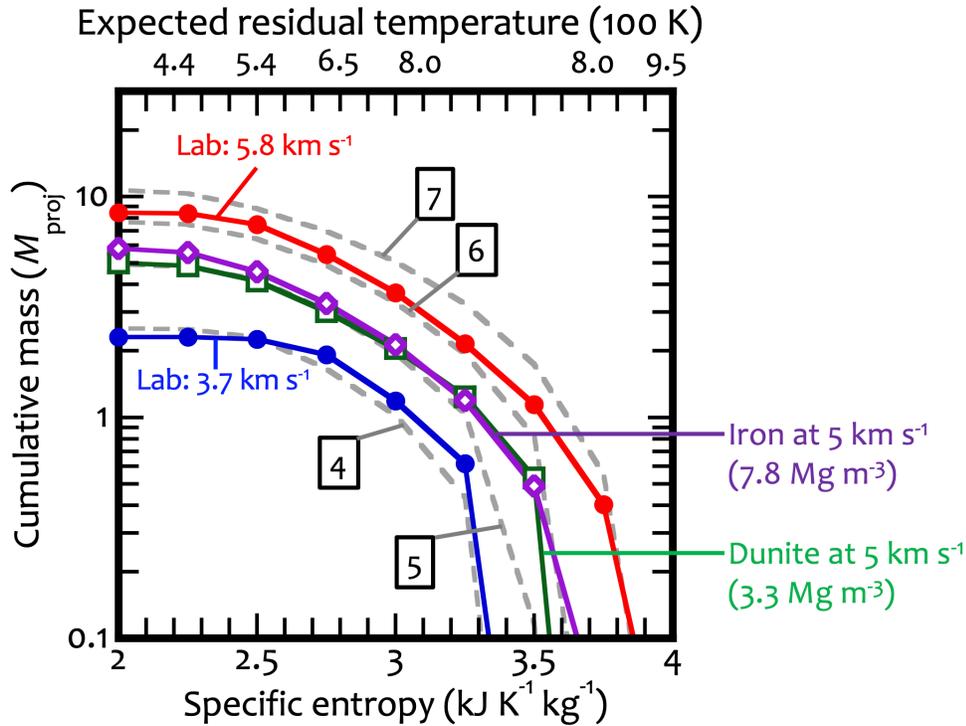

**Supplementary Figure S17.** The same as Fig. 4b in the main text, except that the results for dunite and iron impactors are also shown.

**Supplementary Table S9.** The model input parameters pertaining to a rocky impactor. The values are the same as used in Kurosawa and Genda (2018)[11].

| | |
|---|---|
| EOS model | ANEOS dunite[a] |
| Strength model | Rock[b] |
| Poisson's ratio | 0.25[c] |
| Melting temperature (K) | 1373[c] |
| Thermal softening coefficient | 1.1[c] |
| Simon parameter A (GPa) | 1.52[c] |
| Simon parameter C | 4.05[c] |
| Cohesion (undamaged) (MPa), $Y_{coh,i}$ | 10[d] |
| Cohesion (damaged) (kPa), $Y_{coh}$ | 10[d] |
| Internal friction (undamaged), $\mu_{int}$ | 1.2[d] |
| Internal friction (damaged), $\mu_{dam}$ | $10^{-4}$ to 0.6[e] |
| Limiting strength (GPa), $Y_{limit}$ | 3.5[d] |
| Minimum failure strain | $10^{-4}$ [d] |
| Constant for the damage model | $10^{-11}$ [d] |
| Threshold pressure for the damage model (MPa) | 300[d] |

a. Benz et al. (1989)[26].
b. A detailed description of the strength model for rocks can be found in Collins et al. (2004)[10].
c. Johnson et al. (2015)[31].
d. Typical values for minerals were used.
e. The maximum value was taken from Johnson et al. (2015)[31].

**Supplementary Table S10.** The model input parameters pertaining to a metal impactor. The values are basically the same as used in Bowling et al. (2013)[32].

| | |
|---|---|
| EOS model | ANEOS iron[a] |
| Strength model | Johnson–Cook[b] |
| Poisson's ratio | 0.29[c] |
| Melting temperature (K) | 1811[c] |
| Thermal softening coefficient | 1.2[c] |
| Simon parameter A (GPa) | 6.0[c] |
| Simon parameter C | 3.0[c] |
| Johnson–Cook parameter A (MPa) | 175[b] |
| Johnson–Cook parameter B (MPa) | 380[b] |
| Johnson–Cook parameter N | 0.32[b] |
| Johnson–Cook parameter C | 0.06[b] |
| Johnson–Cook parameter M | 0.55[b] |
| Johnson–Cook parameter $T_{ref}$ (K) | 293[b] |

a. Thompson (1990)[29].
b. The parameters for ARMCO iron from Johnson and Cook (1983)[33] were used.
c. Bowling et al. (2013)[32].

**Supplementary Note S10. Applicable limit of the experimental results in terms of impact velocity**

We showed that the impact experiments at high $v_{imp}$ correspond to a catastrophic disruption of a hypothesised Ryugu parent body at 6–7 km s$^{-1}$, as shown in Fig. 4 in the main text (Supplementary Note S7). In this section, we discuss the feasibility of extrapolation to higher $v_{imp}$. Figure S17 shows the results of the Ryugu-simulations at a velocity range between 0.5 and 10 km s$^{-1}$. Heated masses that have $S_{id} < S < S_{cd}$ ($T_{res} = T_d$) and $S > S_{cd}$ ($T_{res} > T_d$) are shown as a function of $v_{imp}$. The collision probability in the main asteroid belt region[30] is also shown. Figure S17 shows that we only investigated the conditions where the heated mass with $S > S_{cd}$ is negligible. It is not clear whether the local energy concentration described in Note S8.4 occurs under the conditions with $S > S_{cd}$ ($T_{res} > T_d$). Consequently, the experimental results should not be extrapolated to higher impact velocities. Nevertheless, such high-speed collisions characterise only <15% of all collisions in the main asteroid belt region.

The cumulative mass of heated materials having the residual temperature after pressure release (Supplementary Note S8.1) of $T_{res} \geq 800$ K at $v_{imp} = 4$ km s$^{-1}$ in the Ryugu–simulation is ~1 $M_p$ (Supplementary Fig. S17). Such relatively low-speed collisions also lead to considerable heating, whereas the residual temperature distribution is not expected to lead to efficient volatile loss. This is quite different from the estimation by Sugita et al. (2019)[34], who analytically estimated that the residual temperature near the impact point does not exceed 500 K at a collision speed of 5 km s$^{-1}$. This difference is possibly due to neglecting the effect of porosity compaction and shear deformation on the temperature increase in the analytical model. Our experiments show that hypervelocity mutual collisions in the main asteroid belt region are not responsible for significant volatile loss, even when additional heating processes are taken into account using modern hydrocodes. The high-temperature layer beneath the crater floor produced by a catastrophic collision on the Ryugu parent body (Fig. 3b in the main text) is expected to experience only partial devolatilisation, as observed in the impact experiments.

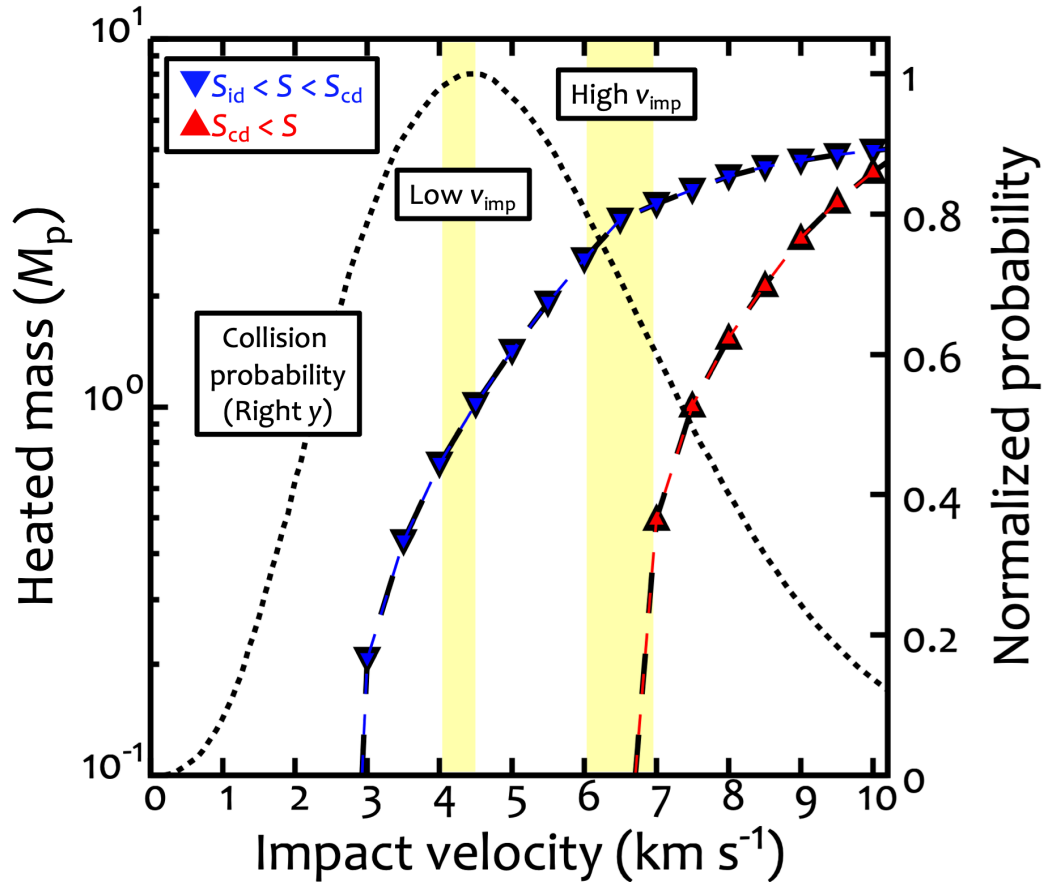

**Supplementary Figure S18.** Heated mass as a function of impact velocity. Results of the Ryugu-simulations with a velocity between 0.5 and 10 km s$^{-1}$ are shown. The heated masses with $S_{id} < S < S_{cd}$ (blue lower triangles) and $S_{cd} < S$ (red upper triangles) were calculated. Yellow shaded regions indicate the velocity ranges investigated with the impact experiments (Supplementary Note S7). The normalised collision probability in the main asteroid belt proposed by Bottke et al. (1994)[30] is also shown on the right $y$-axis.

**Supplementary References**

1. Kurosawa, K., Nagaoka, Y., Senshu, H., Wada, K., Hasegawa, S., Sugita, S. & Matsui, T. Dynamics of hypervelocity jetting during oblique impacts of spherical projectiles investigated via ultrafast imaging. *Journal of Geophysical Research Planets* **120**, 1237–1251 (2015).
2. Kurosawa, K., Moriwaki, R., Komatsu, G., Okamoto, T., Sakuma, H., Yabuta, H. & Matsui, T. Shock vaporization/devolatilization of evaporitic minerals, halite and gypsum, in an open system investigated bbya two-stage light gas gun. *Geophysical Research Letters* **46**, 7258–7267 (2019).
3. Britt, D. T., Cannon, K. M., Donaldson Hanna K., Hogancamp, J., Poch, O., Beck, P., Martin, D., Escrig, J., Bonal, L., & Metzger, P. T. Simulated asteroid materials based on carbonaceous chondrite mineralogies. *Meteoritics & Planetary Science* **54**, 2067–2082 (2019).

4. van Krevelen, D. W. Coal Typology - Physics - Chemistry - Constitution, Amsterdam, the Netherlands : Elsevier Scientific Pub. Co., (1961).
5. Nutting, P. G. Some standard thermal dehydration curves of minerals. *U.S. Geol. Surv. Bull.* **197-E**, 197–216 (1943).
6. Amsden, A., Ruppel, H., & Hirt, C. SALE: A simplified ALE computer program for fluid flow at all speeds. *Los Alamos National Laboratories Report*, LA-8095, 101 pp (1980).
7. Ivanov, B. A., Deniem, D., & Neukum, G. Implementation of dynamic strength models into 2-D hydrocodes: Applications for atmospheric breakup and impact cratering. *Int. J. Impact Eng.* **20**, 411–430 (1997).
8. Wünnemann, K., Collins, G. S., & Melosh, H. J. A strain-based porosity model for use in hydrocode simulations of impacts and implications for transient crater growth in porous targets. *Icarus* **180**, 514–527 (2006).
9. Drucker, D. C. and Prager, W. Soil mechanics and plastic analysis or limit design. *Quarterly of Applied Mathematics* **10**, 157–165 (1952).
10. Collins, G. S., Melosh, H. J., and Ivanov, B. A. Modeling damage and deformation in impact simulations. *Met. And Planet. Sci.,* **39**, 217–231, (2004).
11. Kurosawa, K. and Genda, H. Effects of friction and plastic deformation in shock-comminuted damaged rocks on impact heating. *Geophysical Research Letters* **45**, 620-626, https://doi.org/10.1002/2017GL076285, (2018).
12. Jutzi, M., Michel, P., Benz, W., Richardson, D. C. Fragment properties at the catastrophic disruption threshold: The effect of the parent body's internal structure. Icarus 207, 54–65, (2010).
13. Suetsugu, R., Tanaka, H., Kobayashi, H., and Genda, H. Collisional disruption of planetesimals in the gravity regime with iSALE code: Comparison with SPH code for purely hydrodynamic bodies, *Icarus* **314**, 121–132, (2018).
14. Collins, G. S., Elbeshausen, D., Davison, T. M., Wünnemann, K., Ivanov, B. A., and Melosh, H. J. iSALE-Dellen manual, *Figshare*, https://doi.org/10.6084/m9.figshare.3473690.v2, (2016).
15. Brookshaw, L. An Equation of state for serpentine, *Tech. Rep., working paper series* SC-MC-9813 (Queensland: Faculty of Sciences, University of Southern Queensland) (1998).
16. Davison, T. M., Collins, G. S., and Bland, P. A. Mesoscale modeling of impact compaction of primitive solar system solids. The Astrophysical Journal 821:68, (2016).
17. Kurosawa, K., Genda, H., Azuma, S., & Okazaki, K. The role of post-shock heating by plastic deformation during impact devolatilization of calcite ($CaCO_3$). *Geophysical Research Letters* **48**, e2020GL091130. https://doi.org/10.1029/2020GL091130 (2021).
18. Senshu, H., Kuramoto, K., & Matsui, T. Thermal evolution of a growing Mars. *Journal of Geophysical Research* **107**, 5118, doi:10.1029/2001JE001819.
19. Thompson, S. L. & Lauson, H. S. Improvements in the Chart-D radiation hydrodynamic code III: Revised analytical equation of state, pp. *SC-RR-71 0714* **119** pp., Sandia Laboratories, Albuquerque, NM (1972).
20. Weber, J. N., & Greer, R. T. Dehydration of serpentine: Heat of reaction and reaction kinetics at $P_{H2O}$ = 1 atm. *The American Mineralogist* **50**, 450–464 (1965).
21. Haynes, W. M. CRC handbook of chemistry and physics, (91st ed.). Boca Raton, FL: CRC Press (2010).

22. Chase, M. W., Davies, C. A., Downey, J. R., Frurip, D. R., McDonald, R. A., & Syverud, A. N. JANAF Thermochemical Tables. *Journal of Physical and Chemical Reference Data Supplements*, *14*(1), 1–1856, (1985).
23. Hashimoto, G. L., Abe, Y., & Sugita, S. The chemical composition of the early terrestrial atmosphere: Formation of a reducing atmosphere from CI-like material. *Journal of Geophysical Research* **112**, E05010, doi:10.1029/2006JE002844 (2010).
24. Schaefer, L., & Fegley, B. Jr. Chemistry of atmospheres formed during accretion of the Earth and other terrestrial planets. *Icarus* **208**, 438–448 (2010).
25. Schaefer, L., & Fegley, B. Jr. Redox states of initial atmospheres outgassed on rocky planets and planetesimals. *The Astrophysical Journal* **843**, 120 (2017).
26. Marsh, S. P. *LASL shock Hugoniot data*, 301–302pp., University of California Press, Oakland (1980).
27. Benz, W., Cameron, A. G. W., & Melosh, H. J. The origin of the Moon and the single impact hypothesis III. *Icarus* **81**, 113–131 (1989).
28. Sugita, S., & Schultz, P. H. Interactions between impact-induced vapor clouds and the ambient atmosphere: 2. Theoretical modeling. *Journal of Geophysical Research* **108**, 5052, doi:10.1029/2002JE001960 (2003).
29. King, A. J., Schofield, P. F., & Russell, S. S. Thermal alteration of CM carbonaceous chondrites: Mineralogical changes and metamorphic temperatures. *Geochimica et Cosmochimica Acta* **298**, 167–190 (2021).
30. Thompson, S. L. ANEOS analytic equations of state for shock physics code input manual. *Sandia National Laboratories Report*, 89–2951, UC404 (1990).
31. Bottke, W. F., Nolan, M. C., Greenberg, R., & Kolvoord, R. A. Velocity distributions among colliding asteroids, *Icarus* **107**, 255–268 (1994).
32. Johnson, B. C., Minton, D. A., Melosh, H. J., & Zuber, M. T.. Impact jetting as the origin of chondrules. *Nature* **517**, 339–341, (2015). https://doi.org/10.1038/nature14105
33. Bowling, T. J., Johnson, B. C., Melosh, H. J., Ivanov, B. A., O'Brien, D. P., Gaskell, R., & Marchi, S. Antipodal terrains created by the Rheasilvia basin forming impact on asteroid 4 Vesta. *Journal of Geophysical Research: Planets* **118**, 1821–1834, doi:10.1002/jgre.20123 (2013).
34. Johnson, G. R., & Cook, W. H. A constitutive model and data for metals subjected to large strains, high strain rates and high temperatures. *Seventh International Symposium on Ballistics*, Hague (1983).
35. Sugita, S. et al. The geomorphology, color, and thermal properties of Ryugu: Implications for parent-body processes. *Science* **364**, eaaw0422 (2019).